\PassOptionsToPackage{english}{babel}
\documentclass[reprint,aps,pra,twocolumn,showpacs,amssymb,amsfonts,superscriptaddress,longbibliography]{revtex4-2}  % for review and submission
\usepackage[english]{babel}
\usepackage{graphicx}  % needed for figures
\usepackage{dcolumn}   % needed for some tables
\usepackage{bm}        % for math
\usepackage{bbm}        % for math
\usepackage{amssymb}   % for math
\usepackage{amsmath}
\usepackage{amsthm}
\usepackage{mathtools}
\usepackage[utf8]{inputenc}
\usepackage{siunitx}

\usepackage{changes}

\definecolor{myurlcolor}{rgb}{0,0,0.7}
\definecolor{myrefcolor}{rgb}{0.8,0,0}
\usepackage{hyperref}
\hypersetup{colorlinks, linkcolor=myrefcolor,
citecolor=myurlcolor, urlcolor=myurlcolor}

\usepackage{cleveref}
\crefrangeformat{equation}{#3(#1 #4--#5 #2)#6}

\usepackage{comment}
\usepackage{soul} % for striked out text \st{...}
\newcommand{\ignore}[1]{}
\usepackage{enumitem}
\usepackage{changepage}

\newcommand{\bs}[1]{\boldsymbol{#1}}

\newcommand{\ie}{\textit{i.e. }}				% i.e.
\newcommand{\avg}[1]{\left\langle #1 \right\rangle}		% average
\newcommand{\fnminus}{\ensuremath{f_{N}^-(\mu)}}
\newcommand{\fnplus}{\ensuremath{f_{N}^+(\mu)}}

\begin{document}
\selectlanguage{english}

\title{Multiparameter quantum metrology and mode entanglement \\ with spatially split nonclassical spin states}

\author{Matteo Fadel}
    \email{fadelm@phys.ethz.ch} 
    \affiliation{Department of Physics, ETH Z\"urich, 8093 Z\"urich, Switzerland}
    \affiliation{Department of Physics, University of Basel, Klingelbergstrasse 82, 4056 Basel, Switzerland} 
\author{Benjamin Yadin}
    \affiliation{Naturwissenschaftlich-Technische Fakult\"at, Universit\"at Siegen, Walter-Flex-Straße 3, 57068 Siegen, Germany}
\author{Yuping Mao}
    \affiliation{State Key Laboratory of Precision Spectroscopy, School of Physical and Material Sciences, East China Normal University, Shanghai 200062, China}
\author{Tim Byrnes}
    \affiliation{New York University Shanghai, 1555 Century Ave, Pudong, 200122 Shanghai, China}
    \affiliation{State Key Laboratory of Precision Spectroscopy, School of Physical and Material Sciences, East China Normal University, Shanghai 200062, China}
    \affiliation{NYU-ECNU Institute of Physics at NYU Shanghai, 3663 Zhongshan Road, 200062 Shanghai, China}
    \affiliation{National Institute of Informatics, 2-1-2 Hitotsubashi, Chiyoda-ku, 101-8430 Tokyo, Japan}
    \affiliation{Department of Physics, New York University, New York, 10003 NY, USA}
\author{Manuel Gessner}
    \email{manuel.gessner@icfo.eu}
    \affiliation{ICFO-Institut de Ci\`{e}ncies Fot\`{o}niques, The Barcelona Institute of Science and Technology, Av. Carl Friedrich Gauss 3, 08860, Castelldefels (Barcelona), Spain}

\date{\today}

\begin{abstract}
We identify the multiparameter sensitivity of split nonclassical spin states, such as spin-squeezed and Dicke states spatially distributed into several addressable modes. Analytical expressions for the spin-squeezing matrix of a family of states that are accessible by current atomic experiments reveal the quantum gain in multiparameter metrology, as well as the optimal strategies to maximize the sensitivity. We further study the mode entanglement of these states by deriving a witness for genuine $k$-partite mode entanglement from the spin-squeezing matrix. Our results highlight the advantage of mode entanglement for distributed sensing, and outline optimal protocols for multiparameter estimation with nonclassical spatially-distributed spin ensembles.
\end{abstract}

  \maketitle

\begin{figure}[h]
\includegraphics[width=.41\textwidth]{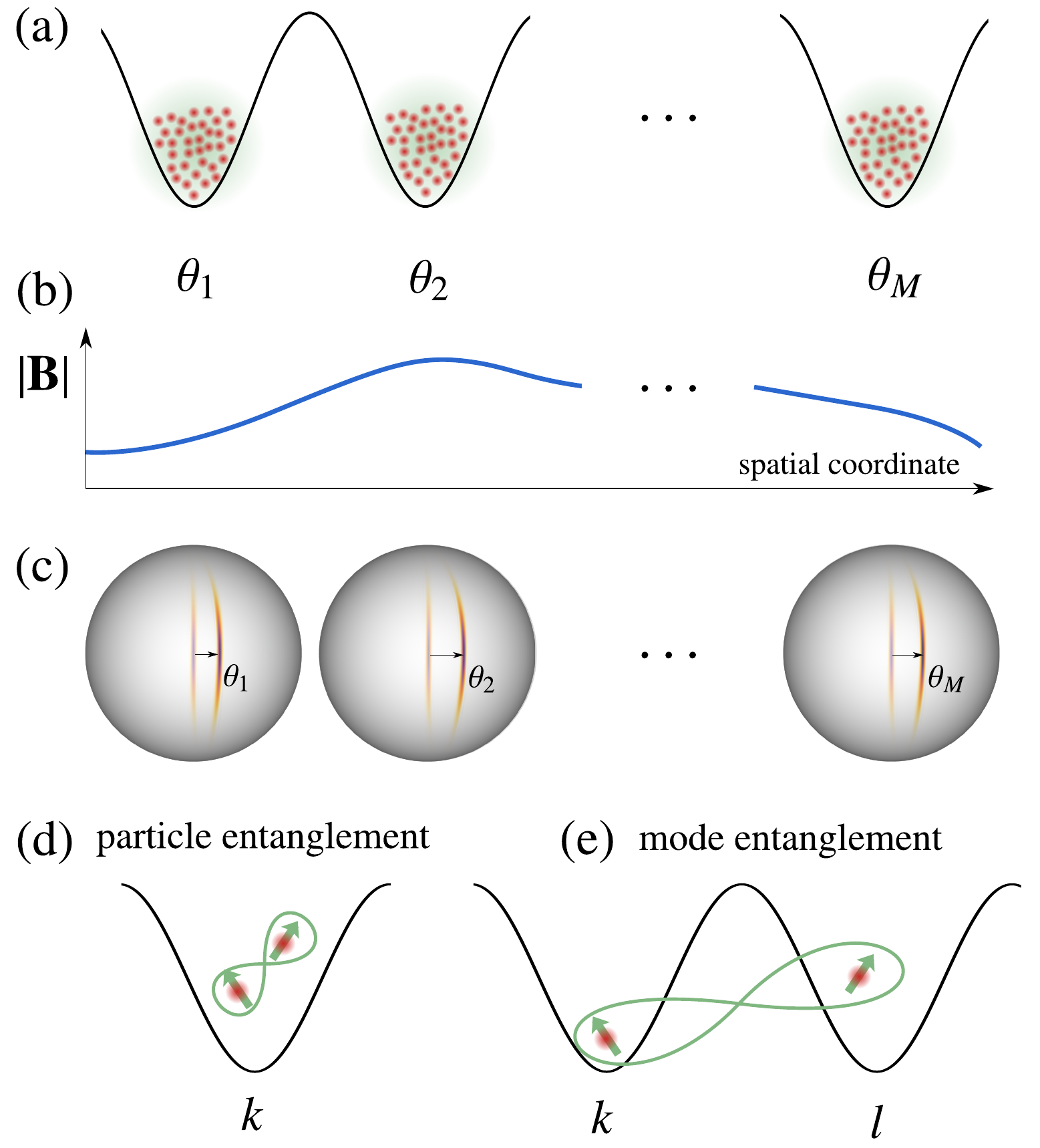}
\caption{\textbf{Multiparameter estimation with a spatially distributed nonclassical spin ensemble}. Each localized spin ensemble occupies a different spatial mode $k=1,\dots,M$ (a) and is subject to a different local electromagnetic field strength (b). The spins therefore experience a different phase shift $\theta_k$ in each mode (c). Strategies to improve the collective measurement sensitivity consist in particle entanglement (d), \ie the entanglement among two spins confined to the same mode $k$, and mode entanglement (e), \ie spin entanglement that is shared between spins in different modes $k\neq l$.}
\label{fig:intro}
\end{figure}

\section{Introduction}
Quantum metrology makes use of non-classical quantum states to enhance measurement precision~\cite{HelstromBook, BraunsteinCaves,Paris2009,GiovannettiNatPhoton2011,TothJPA2014,PezzeRMP2018}. The estimation of a single parameter, e.g., a phase shift in an atomic clock or interferometer, can be made more precise if the atomic spins are prepared in entangled superposition states that have lower quantum fluctuations than classical states. Recently, these ideas have been extended to the problem of multiparameter estimation, where a collective quantum enhancement from a simultaneous estimation of several parameters can be achieved~\cite{HumphreysPRL2013,ProctorPRL2018,GePRL2018,GessnerPRL2018,PolinoOPTICA2019,AlbarelliPLA2020,GoldbergPRL2021,You17}. While the sensitivity limits for general multiparameter scenarios are hard to determine due to the non-commutativity of the observables that provide maximal information on different parameters, this problem can be avoided when all parameters are encoded locally (i.e., the parameter-encoding Hamiltonians commute with each other)~\cite{Matsumoto2002,PezzePRL2017}. In this case, sometimes also called ``distributed sensing'', the collective quantum enhancement can be traced back to the entanglement between the modes where the parameters are encoded~\cite{GessnerPRL2018}. Entanglement in addressable modes can be generated by distributing an ensemble of atomic spins into $M$ spatial modes. This technique has been studied recently both experimentally~\cite{FadelSCIENCE2018,KunkelSCIENCE2018,LangeSCIENCE2018} and theoretically~\cite{Kajtoch18,JingNJP2019,FadelPRA2020} for the case of split spin-squeezed ensembles that can be generated by a nonlinear (one-axis twisting) evolution~\cite{KitagawaUedaPRA1993}.

\clearpage

For single-parameter estimation, the sensitivity gain and the spin entanglement of spin-squeezed states is efficiently captured by the Wineland spin-squeezing parameter~\cite{WinelandPRA1992}. The generalization of this concept to a spin-squeezing matrix quantifies the metrologically relevant quantum fluctuations in the context of multiparameter quantum metrology~\cite{GessnerNATCOMMUN2020}. 

In this article, we identify the multiparameter squeezing matrix of nonclassical spin states split into multiple addressable modes, that are routinely prepared in existing platforms with atomic ensembles, such as, e.g., Bose-Einstein condensates (BECs). We provide exact analytical expressions for the spin-squeezing matrix of spin-squeezed states that are distributed over multiple spatial modes. We distinguish between deterministic and beam-splitter-like distributions of atoms that differ in their partition noise. Furthermore, we introduce a metrological witness for entanglement depth and use it to identify the number of entangled modes from the spin-squeezing matrix. To gauge the ability of the squeezing matrix to describe the full multiparameter sensitivity, we compare to the quantum Fisher matrix. Finally, we discuss possible paths towards a generalization of the spin-squeezing matrix to measurements of nonlinear spin observables and apply it to split Dicke states, whose quantum fluctuations cannot be described by the squeezing of linear spin observables.

\section{Multiparameter sensitivity and spin squeezing matrix}\label{sec:theory}

Assume that a set of $M$ parameters $\bs{\theta}=(\theta_1,\dots,\theta_M)^T$, with $k=1,\dots,M$, is encoded into $M$ spatially separated modes by local rotations. These parameters could, for instance, represent an electromagnetic field at different positions, see Fig.~\ref{fig:intro}. Each rotation is expressed in terms of local collective spin operators $\hat{J}_{\alpha,k} = \sum_{i=1}^{N_k}\hat{\sigma}^{(i)}_{\alpha,k}/2$, where $\hat{\sigma}^{(i)}_{\alpha,k}$ are the Pauli matrices $\alpha=x,y,z$ for the $i$th atom, and $N_k$ is the number of two-level atoms in mode $k$, such that $N=\sum_kN_k$. We consider a parameter-imprinting evolution
\begin{align}\label{eq:unitary}
\hat{U}(\bs{\theta})=\exp\left(-i\sum_{k=1}^M\hat{J}_{\mathbf{r}_k,k}\theta_k\right) \;,
\end{align}
transforming an initial quantum state $\hat{\rho}$ into $\hat{\rho}(\bs{\theta})=\hat{U}(\bs{\theta})\hat{\rho}\hat{U}(\bs{\theta})^{\dagger}$, where $\hat{J}_{\mathbf{r}_k,k}=\mathbf{r}_k^T\hat{\mathbf{J}}_{k}$, $\mathbf{r}_k=(r_{x,k},r_{y,k},r_{z,k})^T$ and $\hat{\mathbf{J}}_k=(\hat{J}_{x,k},\hat{J}_{y,k},\hat{J}_{z,k})^T$ for $k=1,..., M$.

In order to estimate the parameters $\theta_k$, we consider the simultaneous measurement of a vector of local observables $\hat{\mathbf{J}}_{\mathbf{s}}=(\hat{J}_{\mathbf{s}_1,1},\dots,\hat{J}_{\mathbf{s}_M,M})^T$. A straightforward way to construct estimators $\theta_{\mathrm{est},k}$ for all parameters $\theta_k$ is to compare the sample average of repeated measurements of $\hat{J}_{\mathbf{s}}$ with its mean value, which is known from calibration. In the central limit, i.e., after $\eta\gg 1$ repetitions, we obtain a multiparameter estimation error of~\cite{GessnerNATCOMMUN2020}
\begin{align}\label{eq:multisensitivity}
\bs{\Sigma}=(\eta \mathbf{M}[\hat{\rho},\hat{\mathbf{J}}_{\mathbf{r}},\hat{\mathbf{J}}_{\mathbf{s}}])^{-1},
\end{align}
where $\bs{\Sigma}_{kl}=\mathrm{Cov}(\theta_{\mathrm{est},k},\theta_{\mathrm{est},l})$ is the estimator covariance matrix, and
\begin{align}\label{eq:momentmatrix}
\mathbf{M}[\hat{\rho},\hat{\mathbf{J}}_{\mathbf{r}},\hat{\mathbf{J}}_{\mathbf{s}}]=\mathbf{C}[\hat{\rho},\hat{\mathbf{J}}_{\mathbf{r}},\hat{\mathbf{J}}_{\mathbf{s}}] \boldsymbol{\Gamma}[\hat{\rho},\hat{\mathbf{J}}_{\mathbf{s}}]^{-1} 
\mathbf{C}[\hat{\rho},\hat{\mathbf{J}}_{\mathbf{r}},\hat{\mathbf{J}}_{\mathbf{s}}]^T 
\end{align}
is the moment matrix. The latter contains the inverse of the covariance matrix $\bs{\Gamma}[\hat{\rho},\hat{\mathbf{J}}_{\mathbf{s}}]_{kl}=\frac{1}{2}(\langle\hat{J}_{\mathbf{s}_k,k}\hat{J}_{\mathbf{s}_l,l}\rangle_{\hat{\rho}}+\langle\hat{J}_{\mathbf{s}_l,l}\hat{J}_{\mathbf{s}_k,k}\rangle_{\hat{\rho}})-\langle\hat{J}_{\mathbf{s}_k,k}\rangle_{\hat{\rho}}\langle\hat{J}_{\mathbf{s}_l,l}\rangle_{\hat{\rho}}$, and the commutator matrix $(\mathbf{C}[\hat{\rho},\hat{\mathbf{J}}_{\mathbf{r}},\hat{\mathbf{J}}_{\mathbf{s}}])_{kl}=-i\langle[\hat{J}_{\mathbf{r}_k,k},\hat{J}_{\mathbf{s}_l,l}]\rangle_{\hat{\rho}}=\delta_{kl}\langle\hat{J}_{x,k}\rangle_{\hat{\rho}}$. Throughout this article, we define our reference frame for each mode $k$ such that $\mathbf{r}_k$ and $\mathbf{s}_k$ are orthogonal vectors in the $yz$ plane, while the mean-spin direction defines the $x$ direction.

The matrix $\bs{\Sigma}$ contains information about the estimation error for arbitrary linear combinations $\mathbf{n}^T\bs{\theta}$ of the parameters:
\begin{align}\label{eq:lincomb}
\Delta (\mathbf{n}^T\bs{\theta}_{\rm{est}})^2=\mathbf{n}^T\bs{\Sigma}\mathbf{n} \;.
\end{align}
Therefore, the essential information about multiparameter sensitivity is contained in the moment matrix $\mathbf{M}$.

\subsection{Spin-squeezing matrix}\label{sec:sqzmatrix}
In order to motivate the construction of the spin-squeezing matrix, let us first briefly recall the Wineland \textit{et al.} spin-squeezing parameter that expresses the sensitivity gain of single-parameter measurements. For $M=1$, the expression~(\ref{eq:multisensitivity}) reduces to $(\Delta \theta_{\mathrm{est}})^2=(\Delta \hat{J}_{\mathbf{s}})_{\hat{\rho}}^2/(\mu\langle \hat{J}_{x}\rangle_{\hat{\rho}}^2)$. An optimal classical strategy, i.e., in the absence of quantum entanglement, is given by a coherent spin state~\cite{PezzeRMP2018} and achieves an estimation error
$(\Delta \theta_{\mathrm{est}})_{\mathrm{SN}}^2=(\mu N)^{-1}$ at the so-called shot-noise limit. The entanglement-induced quantum enhancement beyond this classical limit is quantified by the Wineland \textit{et al.} spin-squeezing parameter~\cite{WinelandPRA1992}
\begin{align}\label{eq:singleparameterWineland}
    \xi^2[\rho,\hat{J}_{\mathbf{r}},\hat{J}_{\mathbf{s}}]:=\frac{(\Delta \theta_{\mathrm{est}})^2}{(\Delta \theta_{\mathrm{est}})_{\mathrm{SN}}^2}=\frac{N(\Delta \hat{J}_{\mathbf{s}})_{\hat{\rho}}^2}{\langle \hat{J}_{x}\rangle_{\hat{\rho}}^2}.
\end{align}
Any violation of the shot-noise condition $\xi^2[\rho,\hat{J}_{\mathbf{r}},\hat{J}_{\mathbf{s}}]\geq 1$ witnesses entanglement among the spins~\cite{SorensenNATURE2001,SorensenMolmerPRL2001} and indicates a quantum gain for estimations of the unknown phase parameter $\theta$, generated by $\hat{J}_{\mathbf{r}}$, from the measurement observable $\hat{J}_{\mathbf{s}}$. 

A generalization of this idea leads to the spin-squeezing matrix~\cite{GessnerNATCOMMUN2020}. In the considered scenario, the multiparameter shot-noise limit~\cite{GessnerPRL2018} is given by
\begin{align}
\bs{\Sigma}_{\rm{SN}}=(\eta \mathbf{F}_{\rm{SN}})^{-1} \;,
\end{align}
where $\mathbf{F}_{\rm{SN}}=\mathrm{diag}(N_1,\dots,N_M)$. The estimation error~(\ref{eq:multisensitivity}) is therefore above the shot-noise limit, i.e., $\bs{\Sigma}\leq \bs{\Sigma}_{\rm{SN}}$ when 
\begin{align}\label{eq:MSN}
\mathbf{M}[\hat{\rho},\hat{\mathbf{J}}_{\mathbf{r}},\hat{\mathbf{J}}_{\mathbf{s}}]\leq \mathbf{F}_{\rm{SN}} \;.
\end{align}
For square matrices $\mathbf{A}$ and $\mathbf{B}$, the condition $\mathbf{A}\geq \mathbf{B}$ expresses that $\mathbf{A}-\mathbf{B}$ is a positive semi-definite matrix. We write the condition~(\ref{eq:MSN}) equivalently as~\cite{GessnerNATCOMMUN2020}
\begin{align}\label{eq:sqz_cond}
\bs{\xi}^2[\hat{\rho},\hat{\mathbf{J}}_{\mathbf{r}},\hat{\mathbf{J}}_{\mathbf{s}}]\geq\mathbf{1}_M \;,
\end{align}
where the elements of the $M\times M$ spin-squeezing matrix read
\begin{align}\label{eq:winelandmatrix}
(\bs{\xi}^2[\hat{\rho},\hat{\mathbf{J}}_{\mathbf{r}},\hat{\mathbf{J}}_{\mathbf{s}}])_{kl}=\frac{\sqrt{N_kN_l}\mathrm{Cov}(\hat{J}_{\mathbf{s}_k,k},\hat{J}_{\mathbf{s}_l,l})_{\hat{\rho}}}{\langle\hat{J}_{x,k}\rangle_{\hat{\rho}}\langle\hat{J}_{x,l}\rangle_{\hat{\rho}}} \;.
\end{align}
The single-parameter spin-squeezing coefficient~(\ref{eq:singleparameterWineland}) is recovered for $M=1$. 

In multimode settings, it is possible not only to entangle particles in the same mode (particle entanglement), but also to introduce delocalized entanglement among particles that are distributed into different modes (mode entanglement)~\cite{KilloranPRL2014,GessnerPRL2018, FadelPRA2020, MorrisPRX2020}. It has been realized that mode entanglement is a useful resource for achieving collective quantum enhancements for the estimation of linear combinations of parameters that are distributed over multiple modes~\cite{GessnerPRL2018,GePRL2018}. 

Since the shot-noise limit can only be overcome by particle-entangled states~\cite{GessnerPRL2018}, a violation of the condition~(\ref{eq:sqz_cond}) implies particle entanglement among the spins, but does not reveal the distribution of entanglement across the modes. A variety of entanglement witnesses suitable for the detection of mode entanglement are available~\cite{Horodecki,Raymer,HyllusEisert, GessnerPRA2016, GessnerPRA2017, DGCZ, QinNPJQI2019, VanLoock, HuberReview, FadelPRA2020, GuehneToth, GiovannettiPRA2003, JingNJP2019,Vitagliano21}. However, also the spin-squeezing matrix contains information about the correlations between modes in its off-diagonal entries~\cite{GessnerNATCOMMUN2020}. Below, in Sec.~\ref{sec:MSmatrix}, we show how a small modification to the spin-squeezing matrix can transform it into a quantitative witness for genuine multimode entanglement that is able to identify lower bounds on the number of entangled modes.

The spin-squeezing matrix~(\ref{eq:winelandmatrix}) expresses the multiparameter sensitivity obtained by measurements of the angular momentum observables $\hat{\mathbf{J}}_{\mathbf{s}}$. To gauge the ability of this measurement to extract the full metrological features of the quantum state $\hat{\rho}$ under consideration, we compare to the quantum Fisher matrix $\mathbf{F}_Q[\hat{\rho},\hat{\mathbf{J}}_{\mathbf{r}}]$, which represents an upper bound on multiparameter sensitivity for any measurement strategy. Here, this upper bound can be saturated for a pure probe state, since all generators $\hat{J}_{\mathbf{r}_k,k}$ commute with each other~\cite{Matsumoto2002,PezzePRL2017}. We obtain from the multiparameter quantum Cram\'er-Rao bound that the estimation error from an optimal measurement is above shot noise if $\mathbf{F}_Q[\hat{\rho},\hat{\mathbf{J}}_{\mathbf{r}}]\leq \mathbf{F}_{\rm{SN}}$, or equivalently $\boldsymbol{\chi}^{2}[\hat{\rho},\hat{\mathbf{J}}_{\mathbf{r}}]\geq \mathbf{1}_M$ where
\begin{align}\label{eq:9}
    \boldsymbol{\chi}^{2}[\hat{\rho},\hat{\mathbf{J}}_{\mathbf{r}}]&=\mathbf{F}_{\rm{SN}}^{\frac{1}{2}}\mathbf{F}_Q[\hat{\rho},\hat{\mathbf{J}}_{\mathbf{r}}]^{-1}\mathbf{F}_{\rm{SN}}^{\frac{1}{2}}.
\end{align}
and $\mathbf{F}_Q[\hat{\rho},\hat{\mathbf{J}}_{\mathbf{r}}]$ is the quantum Fisher matrix. The moment-based approach gives rise to a lower bound to the  sensitivity of an optimal measurement, i.e., $\mathbf{M}[\hat{\rho},\hat{\mathbf{J}}_{\mathbf{r}},\hat{\mathbf{J}}_{\mathbf{s}}]\leq \mathbf{F}_Q[\hat{\rho},\hat{\mathbf{J}}_{\mathbf{r}}]$, implying that $\bs{\xi}^2[\hat{\rho},\hat{\mathbf{J}}_{\mathbf{r}},\hat{\mathbf{J}}_{\mathbf{s}}]\geq \boldsymbol{\chi}^{2}[\hat{\rho},\hat{\mathbf{J}}_{\mathbf{r}}] $. We hence obtain the following hierarchy of conditions
\begin{align}
    \bs{\xi}^2[\hat{\rho},\hat{\mathbf{J}}_{\mathbf{r}},\hat{\mathbf{J}}_{\mathbf{s}}]\geq \boldsymbol{\chi}^{2}[\hat{\rho},\hat{\mathbf{J}}_{\mathbf{r}}] \geq \mathbf{1}_M,
\end{align}
where the first inequality holds for arbitrary states $\hat{\rho}$, and the second inequality is valid for for shot-noise-limited multiparameter measurements, i.e., particle-separable states $\hat{\rho}$. The strongest condition to check these matrix inequalities is obtained by comparing the respective minimal eigenvalues, i.e.,
\begin{align}\label{eq:hierarchyW}
    \lambda_{\min}(\bs{\xi}^2[\hat{\rho},\hat{\mathbf{J}}_{\mathbf{r}},\hat{\mathbf{J}}_{\mathbf{s}}])\geq \frac{1}{\lambda_{\max}( \boldsymbol{\chi}^{-2}[\hat{\rho},\hat{\mathbf{J}}_{\mathbf{r}}])} \geq 1,
\end{align}
where we used $\lambda_{\min}( \boldsymbol{\chi}^{2}[\hat{\rho},\hat{\mathbf{J}}_{\mathbf{r}}])=\lambda_{\max}( \boldsymbol{\chi}^{-2}[\hat{\rho},\hat{\mathbf{J}}_{\mathbf{r}}])^{-1}$. We refer to $\lambda_{\min}(\bs{\xi}^2[\hat{\rho},\hat{\mathbf{J}}_{\mathbf{r}},\hat{\mathbf{J}}_{\mathbf{s}}])$ as the the \textit{collective squeezing} as it corresponds to the squeezing that can be achieved by the state $\hat{\rho}$ for the estimation of an optimal linear combination of parameters, which in turn is identified by the associated eigenvector [recall Eq.~(\ref{eq:lincomb})]. The hierarchy~(\ref{eq:hierarchyW}) provides us with two pieces of information about multiparameter squeezing. First, a violation of the condition $\lambda_{\min}(\bs{\xi}^2[\hat{\rho},\hat{\mathbf{J}}_{\mathbf{r}},\hat{\mathbf{J}}_{\mathbf{s}}])\geq 1$ identifies a quantum sensitivity enhancement achieved by squeezing, and larger violations imply stronger quantum gains. Second, the difference between $\lambda_{\min}(\bs{\xi}^2[\hat{\rho},\hat{\mathbf{J}}_{\mathbf{r}},\hat{\mathbf{J}}_{\mathbf{s}}])$ and $\lambda_{\max}( \boldsymbol{\chi}^{-2}[\hat{\rho},\hat{\mathbf{J}}_{\mathbf{r}}])^{-1}$ quantifies the metrological quality of the chosen measurement observables $\hat{\mathbf{J}}_{\mathbf{s}}$, i.e., their ability to extract the full sensitivity from the given quantum state. For pure states $\hat{\Psi}=|\Psi\rangle\langle\Psi|$ we can use $\mathbf{F}_Q[\hat{\Psi},\hat{\mathbf{J}}_{\mathbf{r}}]=4\boldsymbol{\Gamma}[\hat{\Psi},\hat{\mathbf{J}}_{\mathbf{r}}]$ to obtain the explicit expression
\begin{align}\label{eq:QFIW}
(\boldsymbol{\chi}^{-2}[\hat{\Psi},\hat{\mathbf{J}}_{\mathbf{r}}])_{kl}=4\frac{\mathrm{Cov}(\hat{J}_{\mathbf{r}_k,k},\hat{J}_{\mathbf{r}_l,l})_{\hat{\Psi}}}{\sqrt{N_kN_l}}.
\end{align}

\subsection{Spin-squeezing matrix for mode entanglement}\label{sec:MSmatrix}

To derive a criterion for mode-separability, we compare the multiparameter sensitivity to the limit achievable by mode-separable states, given by~\cite{GessnerPRL2018}
\begin{align}
\mathbf{M}[\hat{\rho},\hat{\mathbf{J}}_{\mathbf{r}},\hat{\mathbf{J}}_{\mathbf{s}}]\leq \mathbf{F}_{\mathrm{MS}}[\hat{\rho},\hat{\mathbf{J}}_{\mathbf{r}}] \;,
\end{align}
where 
\begin{align}
\mathbf{F}_{\mathrm{MS}}[\hat{\rho},\hat{\mathbf{J}}_{\mathbf{r}}]=4\mathrm{diag}((\Delta\hat{J}_{\mathbf{r}_1,1})_{\hat{\rho}}^2 ,\dots,(\Delta\hat{J}_{\mathbf{r}_M,M})_{\hat{\rho}}^2) \;.
\end{align}
Following the procedure of the preceding Section, we are able to express this condition for mode separability equivalently as
\begin{align}\label{eq:sqz_condMS}
\bs{\xi}_{\mathrm{MS}}^2[\hat{\rho},\hat{\mathbf{J}}_{\mathbf{r}},\hat{\mathbf{J}}_{\mathbf{s}}]\geq \mathbf{1}_M \;,
\end{align}
where
\begin{align}\label{eq:MSmatrix}
(\bs{\xi}_{\mathrm{MS}}^2[\hat{\rho},\hat{\mathbf{J}}_{\mathbf{r}},\hat{\mathbf{J}}_{\mathbf{s}}])_{kl}=\frac{4(\Delta\hat{J}_{\mathbf{r}_k,k})_{\hat{\rho}}(\Delta\hat{J}_{\mathbf{r}_l,l})_{\hat{\rho}}\mathrm{Cov}(\hat{J}_{\mathbf{s}_k,k},\hat{J}_{\mathbf{s}_l,l})_{\hat{\rho}}}{\langle\hat{J}_{x,k}\rangle_{\hat{\rho}}\langle\hat{J}_{x,l}\rangle_{\hat{\rho}}}
\end{align}
is the modified spin-squeezing matrix for mode separability.

As we demonstrate in Appendix~\ref{ap:k-prod-witness}, this construction can be generalized even further to reveal genuine multipartite entanglement among groups of at least $k$ modes. A pure state is called $k$-producible if it can be written as $|\Psi_{k-\mathrm{prod}}\rangle=\bigotimes_{\alpha=1}^b|\psi_{\alpha}\rangle$ and each $|\psi_{\alpha}\rangle$ is an arbitrary quantum state for not more than $k$ parties. A density matrix is $k$-producible if it can be written as a convex linear combination of arbitrary $k$-producible pure states. It is possible to prove (see Appendix~\ref{ap:k-prod-witness}) that any $k$-producible state of modes must satisfy
\begin{align}\label{eq:ksepMS}
\bs{\xi}_{\mathrm{MS}}^2[\hat{\rho},\hat{\mathbf{J}}_{\mathbf{r}},\hat{\mathbf{J}}_{\mathbf{s}}]\geq \frac{1}{k}\mathbf{1}_M.
\end{align}
This inequality is violated if and only if the smallest eigenvalue of the matrix $\bs{\xi}_{\mathrm{MS}}^2[\hat{\rho},\hat{\mathbf{J}}_{\mathbf{r}},\hat{\mathbf{J}}_{\mathbf{s}}]$ is smaller than $1/k$.

Similarly as before, we may compare this criterion to an analogous construction based on the quantum Fisher matrix to gauge the quality of the Gaussian characterization~(\ref{eq:MSmatrix}) of the state's entanglement properties. States that are $k$-producible satisfy $\mathbf{F}_Q[\hat{\rho}_{k-\mathrm{prod}},\hat{\mathbf{J}}_{\mathbf{r}}]\leq k \mathbf{F}_{\mathrm{MS}}[\hat{\rho}_{k-\mathrm{prod}},\hat{\mathbf{J}}_{\mathbf{r}}]$. Following the steps of Eqs.~(\ref{eq:9})--(\ref{eq:hierarchyW}) analogously, we obtain the hierarchy
\begin{align}\label{eq:hierarchyMS}
    \lambda_{\min}(\bs{\xi}_{\mathrm{MS}}^2[\hat{\rho},\hat{\mathbf{J}}_{\mathbf{r}},\hat{\mathbf{J}}_{\mathbf{s}}])\geq \frac{1}{\lambda_{\max}(\boldsymbol{\chi}_{\mathrm{MS}}^{-2}[\hat{\rho},\hat{\mathbf{J}}_{\mathbf{r}}])}\geq \frac{1}{k},
\end{align}
for any mode $k$-producible state, where $\boldsymbol{\chi}_{\mathrm{MS}}[\hat{\rho},\hat{\mathbf{J}}_{\mathbf{r}}]=\mathbf{F}_{\mathrm{MS}}[\hat{\rho},\hat{\mathbf{J}}_{\mathbf{r}}]^{\frac{1}{2}}\mathbf{F}_Q[\hat{\rho},\hat{\mathbf{J}}_{\mathbf{r}}]^{-1}\mathbf{F}_{\mathrm{MS}}[\hat{\rho},\hat{\mathbf{J}}_{\mathbf{r}}]^{\frac{1}{2}}$ and for a pure state we obtain
\begin{align}\label{eq:QFIMS}
    (\boldsymbol{\chi}^{-2}_{\mathrm{MS}}[\hat{\Psi},\hat{\mathbf{J}}_{\mathbf{r}}])_{kl}=\frac{\mathrm{Cov}(\hat{J}_{\mathbf{r}_k,k},\hat{J}_{\mathbf{r}_l,l})_{\hat{\Psi}}}{(\Delta\hat{J}_{\mathbf{r}_k,k})_{\hat{\Psi}}(\Delta\hat{J}_{\mathbf{r}_l,l})_{\hat{\Psi}}}.
\end{align}

\section{Split squeezed states from one-axis-twisting}\label{sec:SSS}
Squeezing represents the leading strategy to achieve quantum enhancements in quantum metrology experiments, from gravitational wave detectors~\cite{TsePRL2019} to atomic clocks~\cite{PezzeRMP2018}. In recent experiments, atomic squeezed spin states were distributed coherently into several addressable modes~\cite{FadelSCIENCE2018,KunkelSCIENCE2018}. In this Section, we study the potential of this approach for multiparameter measurements, as well as the measurable signatures of mode entanglement, by determining the corresponding spin-squeezing matrices~(\ref{eq:winelandmatrix}) and~(\ref{eq:MSmatrix}) analytically.

Generally, we distinguish between two different experimental procedures to achieve spatially distributed squeezed states. The first procedure was followed in the experiments~\cite{FadelSCIENCE2018,KunkelSCIENCE2018,LangeSCIENCE2018} and consists of preparing a squeezed atomic state in a single spatial mode and then dividing this mode coherently into two or more modes via an operation that can be described as a beam splitter on spatial modes. This leads to a probabilistic distribution of atoms in the modes described by a multinomial distribution. As a consequence, partition noise will be present in the spin statistics. Alternatively, we also consider a second procedure, where the atoms are distributed deterministically over the spatial modes. The squeezed state may then be generated, e.g., by a collective interaction with a cavity~\cite{LerouxPRL2010} that affects all atoms in the same way, independently of their spatial mode. This procedure gives rise to a similar split spin-squeezed state, which, however, is free of partition noise.

\subsection{Split squeezed states with partition noise}\label{sec:SSSPN}
Consider an ensemble of $N$ spin-$1/2$ particles, initially prepared in a coherent spin state polarized along the $x$ direction, i.e., $|N/2\rangle_x$ with $J_x|N/2\rangle_x=N/2|N/2\rangle_x$. An evolution of this state generated by the one-axis twisting (OAT) Hamiltonian $H=\hbar \chi J_z^2$ for a time $t=\mu/(2\chi)$ generates squeezing of the collective spin observables and introduces particle entanglement among the individual spins~\cite{KitagawaUedaPRA1993,PezzeRMP2018} in the state $|\Psi(\mu)\rangle=e^{-iH\mu/(2\chi)}|N/2\rangle_x$. Note that the resulting dynamics is cyclic with period $2\pi$, and therefore we limit our attention to the interval $0\leq\mu<2\pi$. For small nonzero $\mu$, the state $|\Psi(\mu)\rangle$ shows along a direction $\mathbf{s}$ in the $yz$-plane a smaller variance than the spin-coherent state, originating from the entanglement created by the nonlinear evolution, while remaining polarized along the $x$ axis.

In this squeezed spin state, all particles are localized in space and occupy the same external (spatial) mode. By applying a beam-splitter transformation to the external mode, the correlated spins can be distributed into $M$ addressable modes with a ratio determined by the probability distribution $p_1,\dots,p_M$, so that on average $N_k=p_k N$ particles are localized in mode $k$. We denote the resulting $M$-mode state by $|\Psi_{\mathrm{PN}}(\mu)\rangle$ and use the notation $\hat{\Psi}_{\mathrm{PN}}(\mu)=|\Psi_{\mathrm{PN}}(\mu)\rangle\langle\Psi_{\mathrm{PN}}(\mu)|$, where the subscript PN indicates the presence of partition noise. The bipartite ($M=2$) version of this scenario has been analyzed theoretically in Ref.~\cite{JingNJP2019} and experimentally with a BEC in Ref.~\cite{FadelSCIENCE2018}. In these works the focus has been the detection of (mode) entanglement and EPR steering between the two partitions, while here our goal is to characterize their potential for applications in multiparameter quantum metrology and to identify entanglement from the metrological properties.

To obtain the metrological properties for multiparameter sensing of this state, we determine all first and second moments of spin observables in each mode for the state $\hat{\Psi}_{\mathrm{PN}}(\mu)$. The local directions for the  measurement $\mathbf{s}_k$ and the rotation $\mathbf{r}_k$ are chosen as the squeezed and anti-squeezed directions, respectively, corresponding to minimal and maximal eigenvectors of the local $2\times 2$ covariance matrices in the $yz$-plane of each mode. The full expressions for first and second moments along arbitrary directions are provided in Appendix~\ref{app:SSSwithPN}, together with the angle specifying the directions $\mathbf{s}_k$ and $\mathbf{r}_k$ [see Eq.~\eqref{eq:SqAngle}], which turn out to be independent of $k$. We obtain 
\begin{subequations}\label{eq:momentsSSS}
\begin{align}
\langle\hat{J}_{x,k}\rangle&=\frac{N}{2}  p_k \cos ^{N-1}\left(\dfrac{\mu}{2} \right) \;, \\
\mathrm{Cov}(\hat{J}_{\mathbf{s}_k,k},\hat{J}_{\mathbf{s}_l,l})&= p_k p_l \frac{N(N-1)}{4} \fnminus +\delta_{kl}p_k \frac{N}{4} \;, \label{eq:momentsSSScov}\\
    \mathrm{Cov}(\hat{J}_{\mathbf{r}_k,k},\hat{J}_{\mathbf{r}_l,l})&= p_k p_l \frac{N(N-1)}{4} \fnplus +\delta_{kl}p_k \frac{N}{4} \;, \label{eq:momentsSSScovAS}
\end{align}
\end{subequations}
where we defined the functions
\begin{align}  \label{eq:fNdef}
&f_{N}^\pm(\mu) =  \frac{1}{4} - \frac{1}{4}\Bigg(\cos ^{N-2}(\mu) \,\mp \\
& \phantom{AAA} \mp \sqrt{\left(\cos^{N-2}(\mu)-1\right)^2+16 \sin ^2\left(\dfrac{\mu}{2}\right) \cos ^{2 N-4}\left(\dfrac{\mu}{2}\right) } \Bigg) \;. \notag
\end{align}
It is easy to check that $\fnminus\leq 0$ and  $\fnplus\geq 0$.

\subsubsection{Spin-squeezing matrix}
We first note that inserting Eq.~(\ref{eq:momentsSSS}) into Eq.~(\ref{eq:winelandmatrix}) leads to
\begin{align}\label{eq:xi2mat}
\bs{\xi}^2[\hat{\Psi}_{\mathrm{PN}}(\mu),\hat{\mathbf{J}}_{\mathbf{r}},\hat{\mathbf{J}}_{\mathbf{s}}]&=\frac{(N-1)\fnminus}{c_N(\mu)}\mathbf{v}\mathbf{v}^T+\frac{1}{c_N(\mu)}\mathbf{1} \;,
\end{align}
where $\mathbf{1}$ is the $M\times M$ identity matrix, $\mathbf{v}=(\sqrt{p_1},\dots,\sqrt{p_M})^T$ is a unit vector, and we have introduced the short-hand notation $c_N(\mu)=\cos^{2N-2}\left(\mu/2\right)$.
The eigenvalues of this matrix can be easily identified as
\begin{align}
    \lambda_{\min}(\bs{\xi}^2[\hat{\Psi}_{\mathrm{PN}}(\mu),\hat{\mathbf{J}}_{\mathbf{r}},\hat{\mathbf{J}}_{\mathbf{s}}])&=\frac{(N-1)\fnminus+1}{c_N(\mu)},\label{eq:lambdaminPN}\\
    \lambda_{\max}(\bs{\xi}^2[\hat{\Psi}_{\mathrm{PN}}(\mu),\hat{\mathbf{J}}_{\mathbf{r}},\hat{\mathbf{J}}_{\mathbf{s}}])&=\frac{1}{c_N(\mu)},
\end{align}
where $\lambda_{\min}$ is non-degenerate for $\mu>0$ [recall that $\fnminus\leq 0$] with eigenvector $\mathbf{v}$ and $\lambda_{\max}$ is $(M-1)$-fold degenerate and corresponds to the eigenspace orthogonal to $\mathbf{v}$. It is easy to verify that the collective squeezing coincides with the single-parameter spin-squeezing~(\ref{eq:singleparameterWineland}) of the spin ensemble before the splitting: $\lambda_{\min}(\bs{\xi}^2[\hat{\Psi}_{\mathrm{PN}}(\mu),\hat{\mathbf{J}}_{\mathbf{r}},\hat{\mathbf{J}}_{\mathbf{s}}])=\xi^2[\hat{\Psi}(\mu),\hat{J}_{\mathbf{r}},\hat{J}_{\mathbf{s}}]$.

The strongest suppression of quantum noise, i.e., the optimal quantum enhancement, is achieved for the estimation of a linear combination of parameters $\mathbf{v}^T\bs{\theta}$, determined by the minimal eigenvector $\mathbf{v}$. It is important to note that this vector can be manipulated by tailoring optimal states that are maximally sensitive for any fixed linear combination of parameters. To see this, first note that the absolute weight of each parameter is determined by the splitting ratio $p_k$. Second, the sign can be modified by applying local rotations: A $\pi$ rotation around the $x$ axis changes the sign of the $k$-th row and $k$-th column of the covariance matrix and thereby of the spin-squeezing matrix~(\ref{eq:winelandmatrix}). Hence, such a rotation, which can be realized with high fidelity in atomic systems with external light fields, introduces a minus sign in the $k$-th component of the vector $\mathbf{v}$. This allows us to engineer a split-squeezed state that maximizes the quantum gain for an arbitrary linear combination of parameters of the form $\mathbf{v}^T\bs{\theta}=\pm\sqrt{p_1}\theta_1\pm\dots\pm\sqrt{p_M}\theta_M$.

Notice that this linear combination is not necessarily the same one that reaches the highest sensitivity, since the quantum gain in each parameter is normalized by the shot-noise limit which depends on the local number of particles $N_k$. When this number is high, the sensitivity is high even if squeezing is only moderate. In order to directly optimize the sensitivity, we must focus on the moment matrix Eq.~(\ref{eq:momentmatrix}), which relates to multiparameter sensitivity via Eqs.~(\ref{eq:multisensitivity}) and~(\ref{eq:lincomb}).

Our analysis based on the squeezing matrix contains only Gaussian properties of the state, i.e., first and second moments of collective spin observables. We may gauge the ability of these expressions to efficiently capture the properties of these states by comparison with more general functions based on the quantum Fisher matrix, see Eqs.~(\ref{eq:hierarchyW}) and~(\ref{eq:hierarchyMS}). Inserting Eq.~(\ref{eq:momentsSSScovAS}) into Eq.~(\ref{eq:QFIW}), we find
\begin{align}\label{eq:chim2SSSpn}
\boldsymbol{\chi}^{-2}[\hat{\Psi}_{\mathrm{PN}}(\mu),\hat{\mathbf{J}}_{\mathbf{r}}]=(N-1)\fnplus\mathbf{v}\mathbf{v}^T +\mathbf{1}\;,
\end{align}
The matrix~(\ref{eq:chim2SSSpn}) has the $(M-1)$-fold degenerate eigenvalue $1$, and the non-degenerate
\begin{align}\label{eq:lambdamaxQFIPN}
\lambda_{\max}(\boldsymbol{\chi}^{-2}[\hat{\Psi}_{\mathrm{PN}}(\mu),\hat{\mathbf{J}}_{\mathbf{r}}])=(N-1)\fnplus+1 \;.
\end{align}
Note that~(\ref{eq:lambdamaxQFIPN}) coincides with $F_Q/N=4 (\Delta \hat{J}_{\mathbf{r}})^2/N$ for one-axis twisting of a single mode with $N$ particles after time $\mu$ [see Eq.~\eqref{eq:momentsSSScovAS}]. We thus recover a multiparameter version of the well-known result that spin squeezing efficiently captures the metrological features of states that can be considered to a good approximation as Gaussian~\cite{PS09,GessnerPRL2019}, corresponding to the early time scales of the OAT evolution.

\subsubsection{Mode-entanglement spin-squeezing matrix}
To analyze the mode entanglement using the modified squeezing matrix~(\ref{eq:MSmatrix}), we make use of the analytical expression for the anti-squeezed variances of split spin-squeezed ensembles, given in Eq.~(\ref{eq:momentsSSScovAS}) for $k=l$. 
For arbitrary $\{p_k\}_{k=1}^M$, we obtain
\begin{align}
    &\qquad\bs{\xi}_{\mathrm{MS}}^2[\hat{\Psi}_{\mathrm{PN}}(\mu),\hat{\mathbf{J}}_{\mathbf{r}},\hat{\mathbf{J}}_{\mathbf{s}}]\\&= A_N(\mu)\frac{(N-1)\fnminus}{c_N(\mu)}\mathbf{w}\mathbf{w}^T + \mathbf{D} \;,\notag
\end{align}
where $A_N(\mu)=1+(N-1)\fnplus\sum_lp_l^2$ and
\begin{align}
w_k &= \sqrt{\frac{p_k(1+p_k(N-1)\fnplus)}{A_N(\mu)}} \;,\\
  D_{k} &= \frac{1 + p_k (N-1)\fnplus}{c_N(\mu)} \;,
\end{align}
are the elements of the vector $\mathbf{w}=(w_{1},\dots,w_{M})^T$ and the diagonal matrix $\textbf{D}$, respectively.

Strategies to analytically compute the eigenvalues for matrices of this form exist \cite{Goulb}, but are in general cumbersome. For simplicity, we focus on the case of equal splitting ratio, \ie $p_k=1/M$ for all $k=1,\dots,M$. In this case, $w_k$ and $D_{k}$ no longer depend on $k$ and $\mathbf{D}$ is proportional to the identity matrix. We find the non-degenerate minimal eigenvalue
\begin{align}\label{eq:lambdaminMSPN}
    &\qquad\lambda_{\min}(\bs{\xi}_{\mathrm{MS}}^2[\hat{\Psi}_{\mathrm{PN}}(\mu),\hat{\mathbf{J}}_{\mathbf{r}},\hat{\mathbf{J}}_{\mathbf{s}}])\\&=\frac{(N-1)\fnminus+1}{c_N(\mu)}\left(1+\frac{N-1}{M}\fnplus\right) \;,\notag
\end{align}
with eigenvector $\mathbf{e}=(1,\dots,1)^T/\sqrt{M}$. Note that in the limit $M\to\infty$, we recover Eq.~(\ref{eq:lambdaminPN}). Intuitively, in this limit, each mode is populated by not more than a single particle and thus the particle entanglement, which is detected by~(\ref{eq:lambdaminPN}), becomes equivalent to the mode entanglement, detected by~(\ref{eq:lambdaminMSPN}).

\begin{figure*}[tb]
\includegraphics[width=\textwidth]{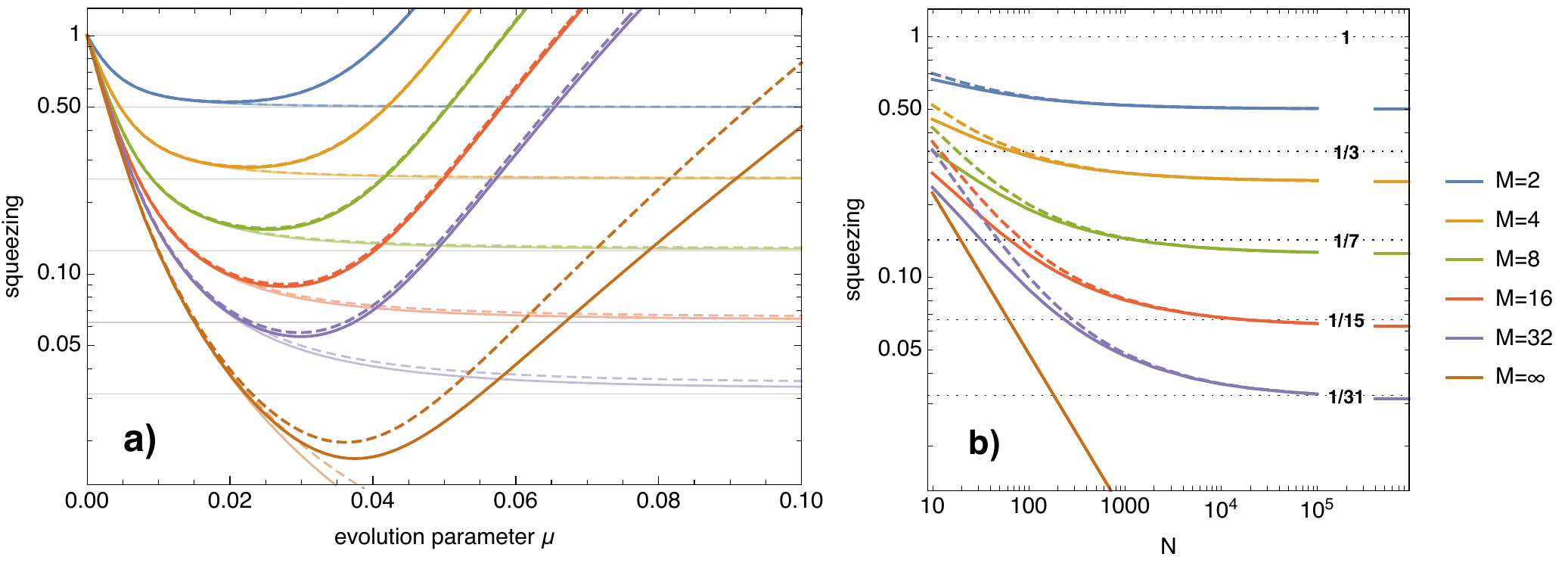}
\caption{\textbf{Mode-separability for spin-squeezed states equally split into $M$ modes.} Dashed and solid curves correspond to splitting with or without partition noise, respectively. \textbf{a)} Minimal eigenvalue of the mode-separability spin-squeezing matrix (thick lines; Eq.~(\ref{eq:lambdaminMSPN}) with partition noise, dashed; Eq.~(\ref{eq:lambdaminMSnPN}) without partition noise, solid) and inverse of the maximum eigenvalue of the mode-separability Fisher matrix (semi-transparent lines; Eq.~(\ref{eq:lmaxQFIMSPN}) with partition noise, dashed; Eq.~(\ref{eq:lmaxQFIMSnoPN}) without partition noise, solid), as a function of the squeezing time $\mu$, and for different $M$, and $N=500$.
Horizontal gray lines corresponds to the asymptotic values $1/M$ for the values of $M$ considered. Note that the optimal squeezing parameter $\mu(M)$ minimizing the eigenvalue of the mode-separability spin-squeezing matrix moves to larger values as $M$ increases. 
\textbf{b)} Minimal eigenvalue of the mode-separability spin-squeezing matrix for the optimal squeezing $\mu(M)$ as a function of $N$, and for different $M$. Note that, as $N\rightarrow\infty$ the eigenvalues tend to $1/M$ (right-most segments). Horizontal dotted lines correspond to the bounds $1/(M-1)$ for the values of $M$ considered; crossing these lines from above implies the detection of genuine multipartite entanglement among all $M$ modes. Note that for a given $M$ there exists a minimum $N$ for which genuine multipartite entanglement can be observed.}
\label{fig:2}
\end{figure*}

The mode entanglement criterion~(\ref{eq:ksepMS}) is shown in Fig.~\ref{fig:2}. We compare the minimal eigenvalue~(\ref{eq:lambdaminMSPN}) to the $k$-separable limit~(\ref{eq:ksepMS}). To observe the strongest possible violation of the separability condition, we optimize the time evolution parameter $\mu$ such that~(\ref{eq:lambdaminMSPN}) takes on its smallest possible value. The optimal squeezing time $\mu_{\mathrm{MS}}$ is generally shorter than the time $\mu_{\mathrm{opt}}$ that optimizes the quantum gain over the shot-noise limit, i.e., the minimal eigenvalue of~(\ref{eq:xi2mat}), whereas both coincide in the limit $M\to\infty$.

Again, we may gauge the quality of our Gaussian spin measurements by comparison with the quantum Fisher matrix via the hierarchy~(\ref{eq:hierarchyMS}). From Eq.~(\ref{eq:momentsSSScovAS}) we can easily obtain the matrix defined in Eq.~(\ref{eq:QFIMS}) in the most general case. In the case of equal splitting ratio, $p_k=1/M$, we obtain
\begin{align}
    \boldsymbol{\chi}^{-2}_{\mathrm{MS}}[\hat{\Psi}_{\mathrm{PN}}(\mu),\hat{\mathbf{J}}_{\mathbf{r}}]&=\frac{\fnplus (N-1)M}{(N-1)\fnplus+M}\mathbf{e}\mathbf{e}^T\notag\\&\quad+\frac{M}{(N-1)\fnplus+M}\mathbf{1} \;.
\end{align}
We find the non-degenerate
\begin{align}\label{eq:lmaxQFIMSPN}
    \lambda_{\max}(\boldsymbol{\chi}^{-2}_{\mathrm{MS}}[\hat{\Psi}_{\mathrm{PN}}(\mu),\hat{\mathbf{J}}_{\mathbf{r}}])=M\frac{(N-1)\fnplus+1}{(N-1)\fnplus+M} \;.
\end{align}
We observe that
\begin{align}
    \lim_{M\to\infty}\lambda_{\max}(\boldsymbol{\chi}^{-2}_{\mathrm{MS}}[\hat{\Psi}_{\mathrm{PN}}(\mu),\hat{\mathbf{J}}_{\mathbf{r}}])=(N-1)\fnplus+1 \;,
\end{align}
hence, in this limit, we recover the maximum eigenvalue~(\ref{eq:lambdamaxQFIPN}) of the matrix~(\ref{eq:chim2SSSpn}). 

The eigenvalues~(\ref{eq:lambdaminMSPN}) and~(\ref{eq:lmaxQFIMSPN}) are plotted in Fig.~\ref{fig:2} as thick and semi-transparent dashed lines, respectively. We visually observe the hierarchy~(\ref{eq:hierarchyMS}) and as the squeezing time $\mu$ increases, we are able to identify genuine multipartite entanglement among larger groups of at least $k$ modes.

\subsection{Split squeezed states without partition noise}
Let us now turn to split squeezed states with a fixed number of particles in each mode. A OAT evolution that acts on all spins collectively, regardless of their spatial mode, generates a split-squeezed state $\hat{\Psi}_{\mathrm{nPN}}(\mu)$ that is free of partition noise. The analytical expressions for the spin expectation values of interest are listed in Appendix~\ref{app:SSSwithoutPN}. As in the previous case, we focus on the spin moments for the optimal directions for spin rotations $\mathbf{r}_k$ and measurements $\mathbf{s}_k$, which correspond to the local squeezed and anti-squeezed spin directions, respectively. These directions are independent of $k$ and coincide with those found previously in the presence of partition noise, since the mode splitting has no impact on the spin state. We obtain
\begin{subequations}\label{eq:momentsSSSNoPN}
\begin{align}
\langle\hat{J}_{x,k}\rangle&=\frac{N_k}{2} \cos ^{N-1}\left(\dfrac{\mu}{2} \right) \;, \\
\mathrm{Cov}(\hat{J}_{\mathbf{s}_k,k},\hat{J}_{\mathbf{s}_l,l})&= \frac{N_k(N_l - \delta_{kl})}{4}\fnminus + 
    \delta_{kl}\frac{N_k}{4} \;, \label{eq:momentsSSScovNoPN}\\
  \mathrm{Cov}(\hat{J}_{\mathbf{r}_k,k},\hat{J}_{\mathbf{r}_l,l}) &=\frac{N_k(N_l - \delta_{kl})}{4}\fnplus + 
    \delta_{kl}\frac{N_k}{4}\:.\label{eq:momentsSSScovNoPNAS}
\end{align}
\end{subequations}

\subsubsection{Spin-squeezing matrix}
Inserting Eq.~(\ref{eq:momentsSSSNoPN}) into Eq.~(\ref{eq:winelandmatrix}) leads to
\begin{align}\label{eq:xi2mat}
\bs{\xi}^2[\hat{\Psi}_{\mathrm{nPN}}(\mu),\hat{\mathbf{J}}_{\mathbf{r}},\hat{\mathbf{J}}_{\mathbf{s}}]&=\frac{N\fnminus}{c_N(\mu)}\mathbf{v}\mathbf{v}^T+\frac{1-\fnminus}{c_N(\mu)}\mathbf{1} \;,
\end{align}
where $\mathbf{v}^T=(\sqrt{N_1/N},\dots,\sqrt{N_M/N})^T$. The eigenvalues read
\begin{align}
    \lambda_{\min}(\bs{\xi}^2[\hat{\Psi}_{\mathrm{nPN}}(\mu),\hat{\mathbf{J}}_{\mathbf{r}},\hat{\mathbf{J}}_{\mathbf{s}}])&=\frac{(N-1)\fnminus+1}{c_N(\mu)} \;,\label{eq:lambdaminnPN}\\
    \lambda_{\max}(\bs{\xi}^2[\hat{\Psi}_{\mathrm{nPN}}(\mu),\hat{\mathbf{J}}_{\mathbf{r}},\hat{\mathbf{J}}_{\mathbf{s}}])&=\frac{1-\fnminus}{c_N(\mu)} \;.
\end{align}
Remarkably, the collective squeezing~(\ref{eq:lambdaminnPN}) coincides with that of~(\ref{eq:lambdaminPN}), indicating that the presence of partition noise does not affect the quantum sensitivity advantage if the squeezing is exploited in an optimal way, i.e., for the linear combination $\mathbf{v}^T\bs{\theta}$ of parameters yielding the largest quantum gain.

For comparison, from Eq.~(\ref{eq:QFIW}), we obtain
\begin{align}
\boldsymbol{\chi}^{-2}[\hat{\Psi}_{\mathrm{nPN}}(\mu),\hat{\mathbf{J}}_{\mathbf{r}}]
     &=N\fnplus\mathbf{v}\mathbf{v}^T+[1-\fnplus]\mathbf{1} \;.
\end{align}
The non-degenerate $\lambda_{\max}(\boldsymbol{\chi}^{-2}[\hat{\Psi}_{\mathrm{nPN}}(\mu),\hat{\mathbf{J}}_{\mathbf{r}}])=(N-1)\fnplus+1$ coincides with the maximum eigenvalue of~(\ref{eq:chim2SSSpn}).

\subsubsection{Mode-entanglement spin squeezing matrix}
For the analysis of mode entanglement using the modified squeezing matrix~(\ref{eq:MSmatrix}), we combine our previous results with the expression~\eqref{eq:momentsSSScovNoPNAS} for the anti-squeezed variances. For arbitrary choices of the $\{N_k\}_{k=1}^M$, we find
\begin{equation}
    \bs{\xi}_{\mathrm{MS}}^2[\hat{\Psi}_{\mathrm{nPN}}(\mu),\hat{\mathbf{J}}_{\mathbf{r}},\hat{\mathbf{J}}_{\mathbf{s}}]= A^{\prime}_N(\mu)\frac{\fnminus}{c_N(\mu)}\mathbf{w}^{\prime}\mathbf{w}^{\prime T} + \mathbf{D}^{\prime} \;,
\end{equation}
where $A^{\prime}_N(\mu)=N+\fnplus\sum_lN_l(N_l-1)$, and the elements of $\mathbf{w}^{\prime}=(w_1^{\prime},\dots,w_M^{\prime})^T$ and the diagonal matrix $\textbf{D}^{\prime}$ are given as
\begin{align}
w^{\prime}_k&=\sqrt{\frac{N_k+N_k(N_k-1)\fnplus}{A^{\prime}_N(\mu)}} \;,\\
D^{\prime}_{k}&= \frac{1-\fnminus}{c_N(\mu)}\left(1+(N_k-1)\fnplus\right) \;,
\end{align}
respectively.

For the special case of equal splitting, i.e., $N_k=N/M$ for all $k$, we obtain the non-degenerate minimal eigenvalue
\begin{align}\label{eq:lambdaminMSnPN}
    &\qquad\lambda_{\min}(\bs{\xi}_{\mathrm{MS}}^2[\hat{\Psi}_{\mathrm{nPN}}(\mu),\hat{\mathbf{J}}_{\mathbf{r}},\hat{\mathbf{J}}_{\mathbf{s}}])\\&=\frac{(N-1)\fnminus+1}{c_N(\mu)}\left(1+\frac{N-M}{M}\fnplus\right) \;.\notag
\end{align}
Comparison with Eq.~(\ref{eq:lambdaminMSPN}) reveals that the presence of partition noise has an effect on the detection of mode entanglement from the spin-squeezing matrix~(\ref{eq:MSmatrix}). A split-squeezed state without partition noise shows a slightly smaller minimal eigenvalue and thus reveals more entanglement at the same nonlinear evolution time $\mu$ according to the witness~(\ref{eq:ksepMS}). A graphical comparison is given in Fig.~\ref{fig:2}, where Eq.~(\ref{eq:lambdaminMSnPN}) is displayed as the thick solid lines.

From Eq.~(\ref{eq:QFIMS}), we obtain for the criterion based on the Fisher information matrix for a uniform splitting ratio
\begin{align}
    \boldsymbol{\chi}^{-2}_{\mathrm{MS}}[\hat{\Psi}_{\mathrm{nPN}}(\mu),\hat{\mathbf{J}}_{\mathbf{r}}]
    &=\frac{\fnplus NM}{(N-M)\fnplus+M}\mathbf{e}\mathbf{e}^T\notag\\&\quad+\frac{M(1-\fnplus)}{ (N-M)\fnplus+M}\mathbf{1} \;.
\end{align}
From this we get
\begin{align}\label{eq:lmaxQFIMSnoPN}
\lambda_{\max}(\boldsymbol{\chi}^{-2}_{\mathrm{MS}}[\hat{\Psi}_{\mathrm{nPN}},\hat{\mathbf{J}}_{\mathbf{r}}])
&=M\frac{\fnplus (N-1)+1}{(N-M)\fnplus+M} \;.
\end{align}
Comparison with~(\ref{eq:lmaxQFIMSPN}) confirms that the influence of partition noise on the mode separability witness remains present when we consider an optimal measurement. The eigenvalues~(\ref{eq:lmaxQFIMSnoPN}) are plotted in Fig.~\ref{fig:2} as semi-transparent solid lines.

\subsection{Sensitivity advantage offered by mode entanglement}\label{sec:modegain}
Let us now compare local (mode separable, Ms) and nonlocal (mode entangled, Me) strategies for the estimation of an arbitrary linear combination of parameters $\mathbf{n}^T\bs{\theta}$. From the results found above, we conclude that--independently of the presence of partition noise--an optimally designed nonlocal strategy can lead to a quantum gain that coincides with the single-parameter spin squeezing coefficient of the initial spin ensemble before splitting, i.e.,
\begin{align}\label{eq:optxi2nonloc}
\xi^2_{\mathrm{Me,opt}}=\mathbf{n}^T\bs{\xi}^2[\hat{\rho}_{\rm Me,opt},\hat{\mathbf{J}}_{\mathbf{r}},\hat{\mathbf{J}}_{\mathbf{s}}]\mathbf{n}=\xi^2[\hat{\rho},\hat{J}_{\mathbf{r}},\hat{J}_{\mathbf{s}}] \;.
\end{align}
For a given linear combination characterized by the coefficients $\mathbf{n}$, this sensitivity is achieved by preparing the optimal nonlocal state $\hat{\rho}_{\rm Me,opt}$ by splitting the maximally squeezed (\ie the state minimizing $\xi^2[\rho,\hat{J}_{\mathbf{r}},\hat{J}_{\mathbf{s}}]$) initial spin ensemble in the state $\hat{\rho}$ with a splitting ratio $p_k=n_k^2$ and then applying local $\pi$-rotations in all modes with negative $n_k$.

To identify the potential advantage of mode entanglement, we compare Eq.~(\ref{eq:optxi2nonloc}) to the quantum gain of an optimal mode-local squeezing strategy with the same average number of particles in each mode. In this case, the spin-squeezing matrix is diagonal, and the multiparameter quantum gain is given by the average of local quantum gains, namely
\begin{align}
\xi^2_{\mathrm{Ms,opt}}=\mathbf{n}^T\bs{\xi}^2[\hat{\rho}_{\rm Ms,opt},\hat{\mathbf{J}}_{\mathbf{r}},\hat{\mathbf{J}}_{\mathbf{s}}]\mathbf{n}=\sum_{k=1}^Mn_k^2\xi^2[\hat{\rho}_k,\hat{J}_{\mathbf{r}_k,k},\hat{J}_{\mathbf{s}_k,k}] \;.
\end{align}
The optimal local strategy consists of maximally squeezing each local spin ensemble, \ie up to the minimum of the local squeezing coefficient $\xi^2[\hat{\rho}_k,\hat{J}_{\mathbf{r}_k,k},\hat{J}_{\mathbf{s}_k,k}]$, respectively.

An advantage of mode entanglement for the estimation of $\mathbf{n}^T\bm{\theta}$ is indicated when the ratio of the respective optimized quantum gains is larger than one, i.e., when
\begin{align}\label{eq:adv}
    \frac{\xi^2_{\mathrm{Ms,opt}}}{\xi^2_{\mathrm{Me,opt}}}>1.
\end{align}

\begin{figure}[tb]
\includegraphics[width=.48\textwidth]{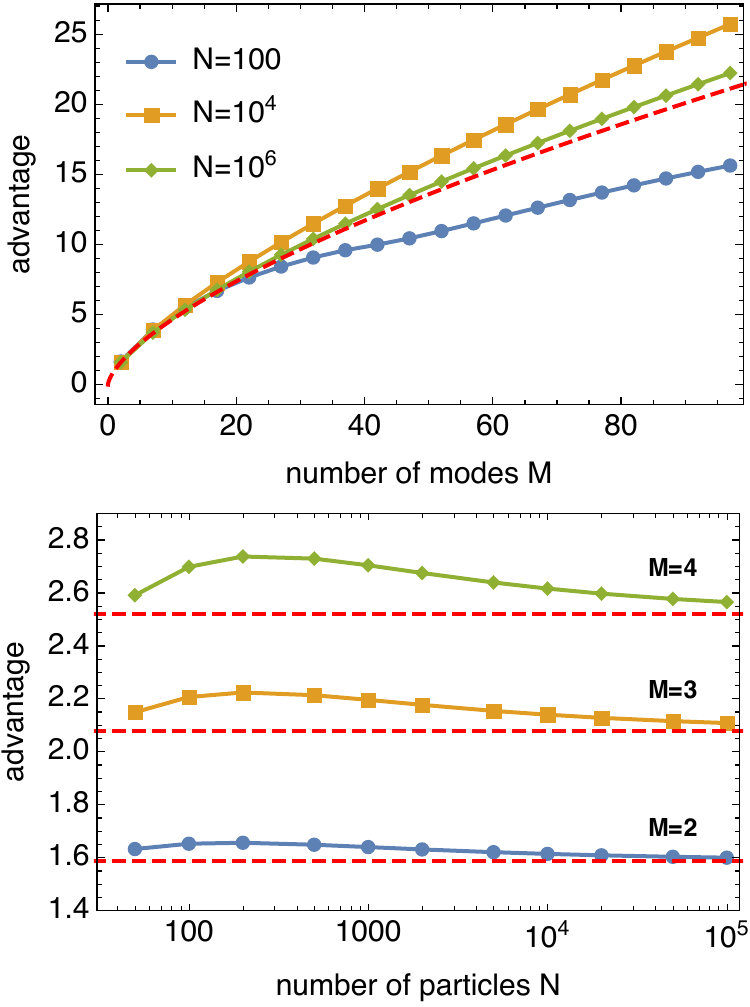}
\caption{\textbf{Advantage of mode entanglement}. Top panel: ratio of local and nonlocal gain over the multiparameter shot-noise limit $\frac{\xi^2_{\mathrm{Ms,opt}}}{\xi^2_{\mathrm{Me,opt}}}$ for an equally weighted linear combination of parameters as a function of the number of modes $M$ and different values of $N$. For the preparation of the optimal nonlocal probe state, the BEC is split equally into $M$ modes after a squeezing evolution up to maximum squeezing. The local strategy consists of optimal local squeezing evolutions of individual BECs whose particle number $N/M$ coincides with the average particle number in each mode of the nonlocal state. The red dashed line represents the analytical prediction~(\ref{eq:anam32}) for $N\rightarrow\infty$. We plot $N=100$ (blue), $N=10^4$ (orange) and $N=10^6$ (green). Bottom panel: Same ratios as before as a function of the total atom number $N$, for splitting into $M=2$ (blue), $M=3$ (orange), $M=4$ (green) modes. The red dashed line represents the analytical prediction~(\ref{eq:anam32}) for $N\rightarrow\infty$.}
\label{fig:gainME}
\end{figure}

For large number of particles $N$, the scaling of this figure of merit can be determined analytically. The single-parameter spin squeezing coefficient for $N$ particles at the optimal squeezing time behaves asymptotically as~\cite{KitagawaUedaPRA1993,SinatraFront2012}
\begin{align}
\xi^2[\hat{\rho},\hat{J}_{\mathbf{r}},\hat{J}_{\mathbf{s}}]\simeq\frac{3^{\frac{2}{3}}}{2} N^{-\frac{2}{3}} 	 \qquad (N\gg 1) \;.
\end{align}
Since the optimal mode-entangled strategy allows us to make use of the collective squeezing of all particles, we obtain $\xi^2_{\mathrm{Ms,opt}}=\frac{3^{\frac{2}{3}}}{2} N^{-\frac{2}{3}}$, whereas in each local mode we only have $p_kN$ particles. We now focus on the case of the estimation of an equally weighted linear combination of parameters, i.e., $|n_k|=1/\sqrt{M}$. The optimal splitting ratio for the nonlocal strategy in this case is also an equally weighted distribution of $N/M$ atoms among all modes. Thus each local spin squeezing parameter yields $\xi^2[\hat{\rho}_k,\hat{J}_{\mathbf{r}_k,k},\hat{J}_{\mathbf{s}_k,k}]=\frac{3^{\frac{2}{3}}}{2} (N/M)^{-\frac{2}{3}}$. Consequently the additional gain provided by mode entanglement is given by
\begin{align}\label{eq:anam32}
\frac{\xi^2_{\mathrm{Ms,opt}}}{\xi^2_{\mathrm{Me,opt}}}=M^{2/3}.
\end{align}
The behavior of the quantum gain at numerically determined optimal squeezing times are compared to the analytical prediction Eq.~(\ref{eq:anam32}) in Fig.~\ref{fig:gainME}. Condition~(\ref{eq:adv}) is fulfilled for arbitrary values of $N$ and $M$, demonstrating the increased quantum gain that is offered by mode-entangled strategies. We  further observe how the asymptotic prediction~(\ref{eq:anam32}), which is shown as red dashed line in both panels, is approached with increasing $N$.

\section{Split Dicke states}\label{sec:STF}
\begin{figure}[tb]
\includegraphics[width=.4\textwidth]{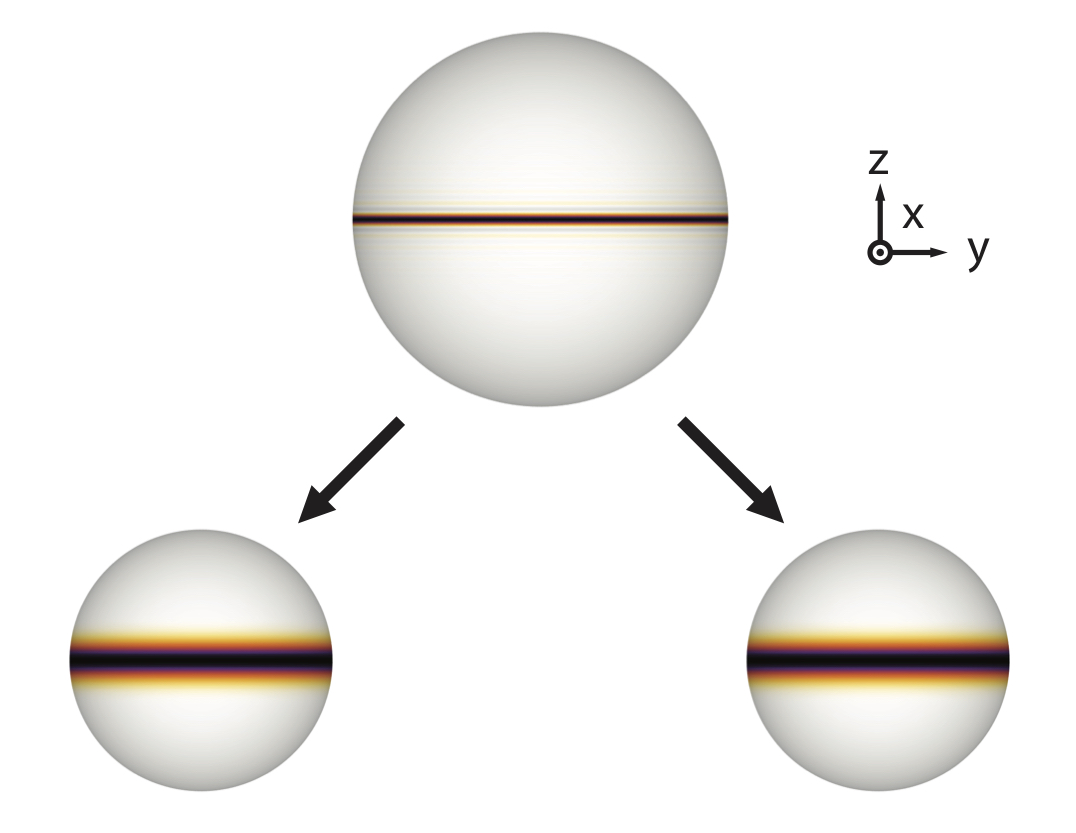}
\caption{\textbf{Split twin-Fock state.} Wigner function of a split twin-Fock state with partition noise in $M=2$ modes, represented on a generalized Bloch sphere of $N$ qubits. The local states after splitting are obtained from the partial trace over the respective other subsystem.}
\label{fig:4}
\end{figure}

In the previous Section we focused on applications with squeezed spin states that are well characterized by averages and variances of collective spin observables. This formalism is, however, no longer suitable for non-Gaussian spin states, such as Dicke states (see Fig.~\ref{fig:4}) that can also be generated experimentally in BECs~\cite{LuckeSCIENCE2011,LuckePRL2014}. For single-parameter measurements, the Wineland spin-squeezing coefficient has been generalized also to nonlinear measurements to account for the fluctuations of non-Gaussian states~\cite{GessnerPRL2019,BaamaraPRL2021}. In Sec.~\ref{sec:nonlinear}, we show how generalized squeezing matrices can be constructed from more general local measurement observables, beyond collective spin components. Then, in Sec.~\ref{sec:sD}, we apply this concept to split Dicke states. We observe that, in contrast to the case of Gaussian squeezed states, local measurements (even of nonlinear operators) are no longer able to capture the state's full multiparameter sensitivity due to the nonlinearity of the optimal observables.

\subsection{Spin-squeezing matrices from nonlinear measurements}\label{sec:nonlinear}
In order to generalize the construction of the spin-squeezing matrix and its variants, we consider the measurement of a vector of local observables $\hat{\mathbf{X}}_{\mathbf{s}}=(\hat{X}_{\mathbf{s}_1,1},\dots,\hat{X}_{\mathbf{s}_M,M})^T$. Here, the observables $\hat{X}_{\mathbf{s}_k,k}$ may contain higher-order moments of the local collective angular moment observables in the mode $k$. The value of the phases $\bs{\theta}$, imprinted as before by a set of local collective spin operators $\hat{\mathbf{J}}_{\mathbf{r}}$, is estimated from the average results using the method of moments~\cite{GessnerNATCOMMUN2020}. We obtain in the central limit ($\eta\gg 1$ repeated measurements) a multiparameter sensitivity of
\begin{align}\label{eq:multisensitivityNL}
\bs{\Sigma}=(\eta \mathbf{M}[\hat{\rho},\hat{\mathbf{J}}_{\mathbf{r}},\hat{\mathbf{X}}_{\mathbf{s}}])^{-1},
\end{align}
where the moment matrix for such a nonlinear measurement is described as
\begin{align}\label{eq:momentmatrixNL}
\mathbf{M}[\hat{\rho},\hat{\mathbf{J}}_{\mathbf{r}},\hat{\mathbf{X}}_{\mathbf{s}}]=\mathbf{C}[\hat{\rho},\hat{\mathbf{J}}_{\mathbf{r}},\hat{\mathbf{X}}_{\mathbf{s}}] \boldsymbol{\Gamma}[\hat{\rho},\hat{\mathbf{X}}_{\mathbf{s}}]^{-1} 
\mathbf{C}[\hat{\rho},\hat{\mathbf{J}}_{\mathbf{r}},\hat{\mathbf{X}}_{\mathbf{s}}]^T.
\end{align}

Since the separability limits are derived from generally valid upper sensitivity limits that depend only on the generators but not on the measurement observables, we can define the squeezing matrix, in direct analogy to the approach presented in Sec.~\ref{sec:sqzmatrix}, as
\begin{align}\label{eq:winelandmatrixNL}
(\bs{\xi}^2[\hat{\rho},\hat{\mathbf{J}}_{\mathbf{r}},\hat{\mathbf{X}}_{\mathbf{s}}])_{kl}=\frac{\sqrt{N_kN_l}\mathrm{Cov}(\hat{X}_{\mathbf{s}_k,k},\hat{X}_{\mathbf{s}_l,l})_{\hat{\rho}}}{-\langle[\hat{J}_{\mathbf{r}_k,k},\hat{X}_{\mathbf{s}_k,k}]\rangle_{\hat{\rho}}\langle[\hat{J}_{\mathbf{r}_l,l},\hat{X}_{\mathbf{s}_l,l}]\rangle_{\hat{\rho}}} \;,
\end{align}
and all particle-separable states must satisfy~\cite{GessnerNATCOMMUN2020}
\begin{align}\label{eq:sqz_condNL}
\bs{\xi}^2[\hat{\rho}_{\mathrm{p-sep}},\hat{\mathbf{J}}_{\mathbf{r}},\hat{\mathbf{X}}_{\mathbf{s}}]\geq\mathbf{1}_M \;,
\end{align}
which is equivalent to shot-noise-limited multiparameter sensitivities.

Following an analogous procedure as in Sec.~\ref{sec:MSmatrix}, we define the mode-separability squeezing matrix as
\begin{align}\label{eq:MSmatrixNL}
&\quad(\bs{\xi}_{\mathrm{MS}}^2[\hat{\rho},\hat{\mathbf{J}}_{\mathbf{r}},\hat{\mathbf{X}}_{\mathbf{s}}])_{kl}\notag\\&=\frac{4(\Delta\hat{J}_{\mathbf{r}_k,k})_{\hat{\rho}}(\Delta\hat{J}_{\mathbf{r}_l,l})_{\hat{\rho}}\mathrm{Cov}(\hat{X}_{\mathbf{s}_k,k},\hat{X}_{\mathbf{s}_l,l})_{\hat{\rho}}}{-\langle[\hat{J}_{\mathbf{r}_k,k},\hat{X}_{\mathbf{s}_k,k}]\rangle_{\hat{\rho}}\langle[\hat{J}_{\mathbf{r}_l,l},\hat{X}_{\mathbf{s}_l,l}]\rangle_{\hat{\rho}}} \;,
\end{align}
i.e., any mode $k$-producible state must satisfy
\begin{align}\label{eq:sqz_condk}
\bs{\xi}_{\mathrm{MS}}^2[\hat{\rho}_{k-\mathrm{prod}},\hat{\mathbf{J}}_{\mathbf{r}},\hat{\mathbf{X}}_{\mathbf{s}}]\geq\frac{1}{k}\mathbf{1}_M.
\end{align}
These definitions hold for arbitrary choices of the local measurement observables $\hat{\mathbf{X}}_{\mathbf{s}}$. Notice also that the definitions~(\ref{eq:QFIW}) and~(\ref{eq:QFIMS}) based on the quantum Fisher matrix are unaffected by this generalization, since they are already independent of the chosen measurement observables by virtue of a systematic optimization.

\subsection{Split Dicke states}
The highly sensitive features of Dicke states~\cite{LuckeSCIENCE2011} can be efficiently captured by a nonlinear spin measurement up to second order. In the following Sec.~\ref{sec:singlemodeDicke} we identify the optimal second-order observable for arbitrary single-mode Dicke states. In Sec.~\ref{sec:sD} we explore the potential of local measurements of this observable for multiparameter metrology with a split Dicke state and identify the limitations of local measurement strategies for multiparameter quantum metrology with non-Gaussian states that contain mode entanglement.

\subsubsection{Single-mode Dicke states}\label{sec:singlemodeDicke}
To identify an optimal second-order measurement observable, we first focus on the estimation of a single parameter using a single-mode Dicke state. Generally, for any set of accessible observables $\hat{\mathbf{A}}$, the maximally achievable sensitivity for estimations of an angle imprinted by the generator $\hat{H}_{\mathbf{r}}=\mathbf{r}\cdot\hat{\mathbf{H}}$ using the method of moments is given by $\mathbf{r}^T\mathbf{M}[\hat{\rho},\hat{\mathbf{H}},\hat{\mathbf{A}}]\mathbf{r}$ where~\cite{GessnerPRL2019}
\begin{align}\label{eq:singlemodemomentmatrixNL}
\mathbf{M}[\hat{\rho},\hat{\mathbf{H}},\hat{\mathbf{A}}]=\mathbf{C}[\hat{\rho},\hat{\mathbf{H}},\hat{\mathbf{A}}] \, \boldsymbol{\Gamma}[\hat{\rho},\hat{\mathbf{A}}]^{-1} \,
\mathbf{C}[\hat{\rho},\hat{\mathbf{H}},\hat{\mathbf{A}}]^T,
\end{align}
and the optimal linear combination within this operator family $\hat{\mathbf{A}}$ achieving this sensitivity is determined as $\hat{X}_{\mathbf{s}}=\mathbf{s}\cdot\hat{\mathbf{X}}$ with~\cite{GessnerPRL2019}
\begin{align}
    \mathbf{s}=\alpha \mathbf{C}[\hat{\rho},\hat{\mathbf{H}},\hat{\mathbf{A}}]\boldsymbol{\Gamma}[\hat{\rho},\hat{\mathbf{A}}]^{-1}\mathbf{r},
\end{align}
and $\alpha\in\mathbb{R}$ is an arbitrary constant.

To capture the nonlinear features of a Dicke state in mode $k$, we add to the set of 3 linear measurement observables $\hat{\mathbf{J}}_k$ all symmetrized operators of second order, i.e., $\{\hat{J}_{\alpha,k},\hat{J}_{\beta,k}\}/2$ with $\alpha,\beta\in\{x,y,z\}$. We obtain a family of 9 operators that can be used to express arbitrary spin observables of second order. We note that symmetrized second-order operators can be extracted by measuring expectation values of $(\hat{J}_{x,k}+\hat{J}_{z,k})^2$, $\hat{J}_{x,k}^2$ and $\hat{J}_{z,k}^2$, using $\{\hat{J}_{x,k},\hat{J}_{z,k}\} = (\hat{J}_{x,k}+\hat{J}_{z,k})^2 - \hat{J}_{x,k}^2 - \hat{J}_{z,k}^2$. 

For the Dicke state $|j,m\rangle$ with $\hat{J}_{z,k}|j,m\rangle=m|j,m\rangle$, considering the family of 9 observables up to second order $\hat{\mathbf{A}}_k$ and 3 first-order generators $\hat{\mathbf{J}}_k$, it is straightforward to verify that the commutator matrix $\mathbf{C}[\hat{\rho},\hat{\mathbf{J}}_k,\hat{\mathbf{X}}_k]$ is zero everywhere except for
\begin{align}
    -i\langle[\hat{J}_{x,k},\hat{J}_{y,k}]\rangle_{|j,m\rangle}&=m,\notag\\
    -i\langle[\hat{J}_x,\frac{1}{2}\{\hat{J}_{y,k},\hat{J}_{z,k}\}]\rangle_{|j,m\rangle}&=-\frac{1}{2}(j(j+1)-3m^2),\notag\\
    -i\langle[\hat{J}_y,\frac{1}{2}\{\hat{J}_{x,k},\hat{J}_{z,k}\}]\rangle_{|j,m\rangle}&=\frac{1}{2}(j(j+1)-3m^2).
\end{align}
This means that we can limit our attention to the family of measurement observables $\hat{\mathbf{X}}_k=(\hat{J}_{x,k},\hat{J}_{y,k},\frac{1}{2}\{\hat{J}_{x,k},\hat{J}_{z,k}\},\frac{1}{2}\{\hat{J}_{y,k},\hat{J}_{z,k}\})^T$. The symmetry of the Dicke states around the $z$ axis further allow us to focus only on rotations generated by $\hat{J}_{x,k}$ and $\hat{J}_{y,k}$. Restricting to the set $\hat{\mathbf{X}}_k$ furthermore removes the singularity of the full $9\times 9$ covariance matrix $\boldsymbol{\Gamma}[|j,m\rangle,\hat{\mathbf{A}}_k]$, and we obtain (see Appendix~\ref{app:DickeSingleMode} for details)
\begin{align}
    \mathbf{M}[|j,m\rangle,\begin{pmatrix}\hat{J}_{x,k}\\\hat{J}_{y,k}\end{pmatrix},\hat{\mathbf{X}}_k]=2(j(j+1)-m^2)\mathbf{1}_2.
\end{align}
Due to the symmetry of Dicke states (see Fig.~\ref{fig:4}), the sensitivity $2(j(j+1)-m^2)$ is independent of the rotation axis $\mathbf{r}_k=(r_{x,k},r_{y,k},0)^T$ in the $xy$-plane. This sensitivity indeed coincides with the quantum Fisher information matrix of Dicke states
\begin{align}\label{eq:QFIDicke}
    \mathbf{F}_Q[|j,m\rangle,\begin{pmatrix}\hat{J}_{x,k}\\\hat{J}_{y,k}\end{pmatrix}]=2(j(j+1)-m^2)\mathbf{1}_2,
\end{align}
thus demonstrating the optimality of the considered measurements. The optimal observable, however, depends on $\mathbf{r}_k$ and reads
\begin{align}\label{eq:optlocalDicke}
    \hat{X}_{\mathrm{opt},k}&=r_{x,k}\left(\frac{1}{2}\{\hat{J}_{y,k},\hat{J}_{z,k}\}-m \hat{J}_{y,k}\right)\notag\\&\quad+r_{y,k}\left(\frac{1}{2}\{\hat{J}_{x,k},\hat{J}_{z,k}\}- m \hat{J}_{x,k}\right).
\end{align}

\subsubsection{Split Dicke states}\label{sec:sD}
We now try to extend these ideas to a multiparameter sensing protocol based on split multimode Dicke states, where in each mode $k$, an optimal local observable is measured, in analogy to the strategy discussed above for split squeezed states. We therefore suppose that each local parameter $\theta_k$ is estimated from the measurement results of the observable $\hat{X}_{\mathbf{s}_k,k}=\mathbf{s}_k\cdot\hat{\mathbf{X}}_k$ with $\mathbf{s}_k=(-mr_{y,k},-mr_{x,k},r_{y,k},r_{x,k})^T$ chosen to match the optimal local measurement observable~(\ref{eq:optlocalDicke}). The rotations are locally generated by $\hat{J}_{\mathbf{r}_k,k}$ around the axis $\mathbf{r}_k=(r_{x,k},r_{y,k},0)^T$.

\begin{figure}[tb]
\includegraphics[width=\columnwidth]{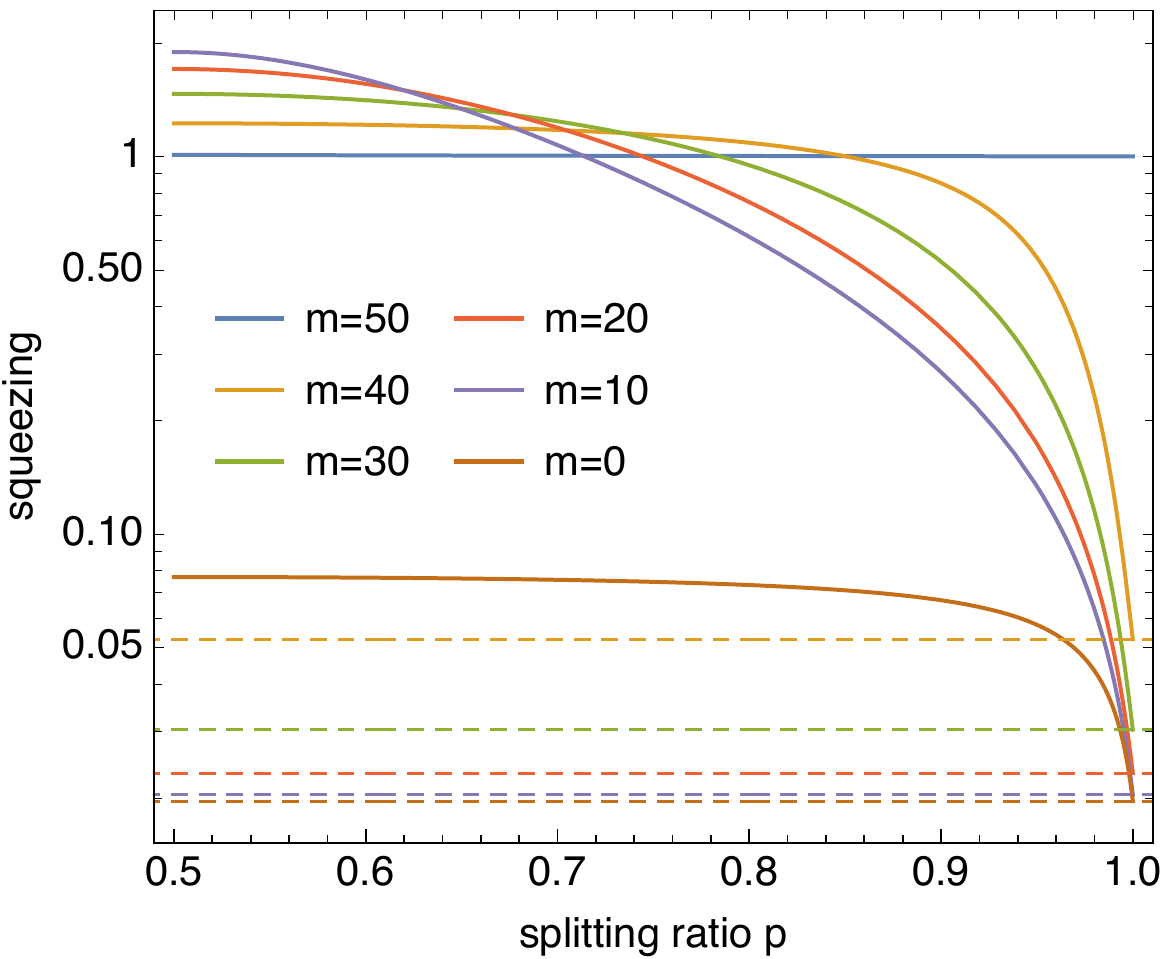}
\caption{\textbf{Sensitivity for Dicke states split into two modes.} For $N=100$ we show the minimal eigenvalue of the spin-squeezing matrix matrix~(\ref{eq:winelandmatrixNL}) as a function of the splitting ratio $p$, and for different $m$.
Horizontal dashed lines correspond to the values expected for the sensitivity of a Dicke state before splitting, which cannot be reached by applying this local, nonlinear measurement strategy to a split Dicke state.}
\label{fig:sDicke2}
\end{figure}

In the following we focus on the relevant case of split Dicke states $|j,m\rangle$ in the presence of partition noise \cite{LangeSCIENCE2018,Vitagliano21}, i.e., splitting is created by a beam splitter operation on the spatial modes, leading to the state $\hat{\Psi}_{j,m,\mathrm{PN}}$. The full analytical expressions for the elements of the relevant covariances and commutators are given in the Appendix~\ref{app:splitDicke}. These allow for a straightforward construction of the spin-squeezing matrices~(\ref{eq:winelandmatrixNL}) and~(\ref{eq:MSmatrixNL}), whose full expressions are rather lengthy and we therefore omit them here. In Fig.~\ref{fig:sDicke2}, the minimal eigenvalue of the squeezing matrix~(\ref{eq:winelandmatrixNL}) is plotted for two-mode split Dicke states as a function of the splitting ratio $p:1-p$ for different values of $m$.

\begin{figure}[tb]
\includegraphics[width=\columnwidth]{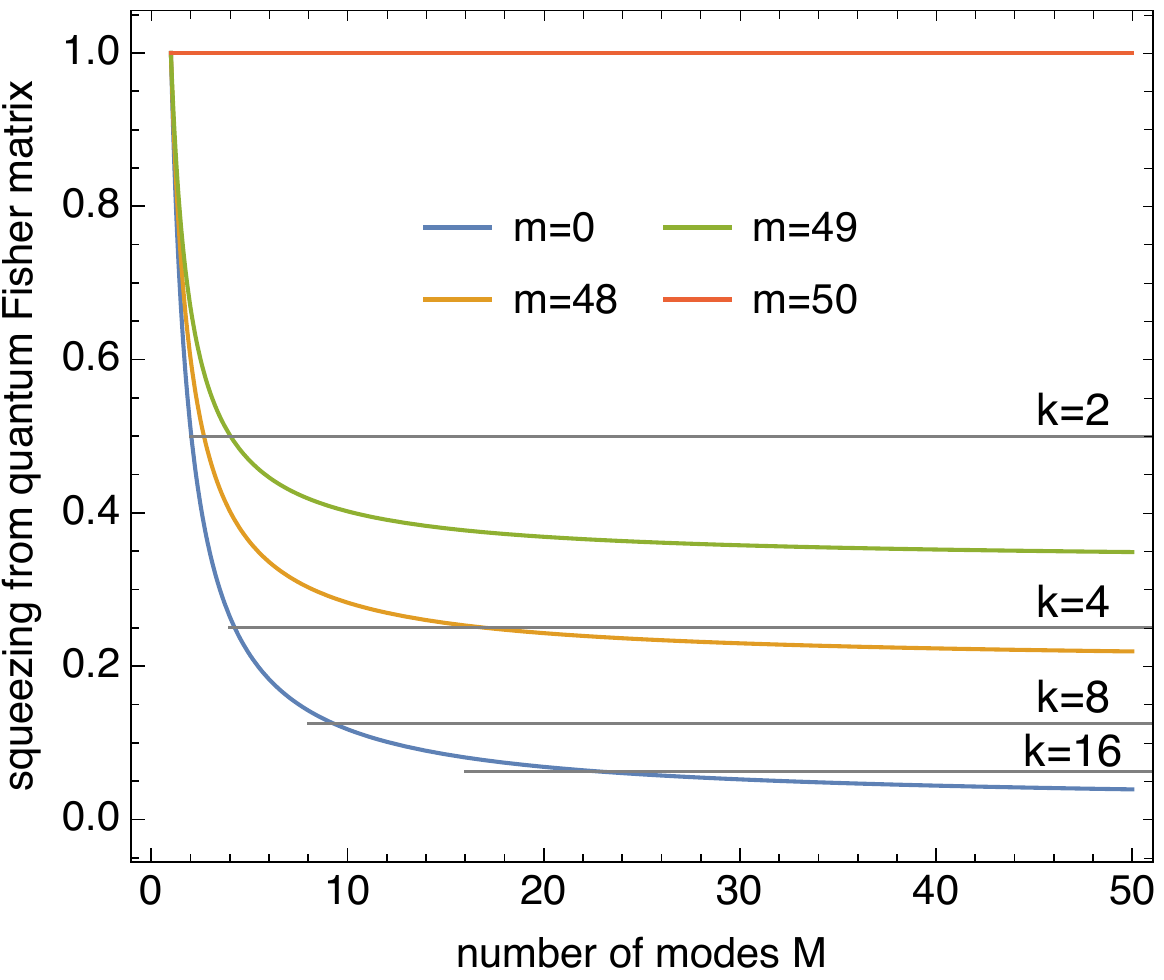}
\caption{\textbf{Mode entanglement of split Dicke states.} We plot the inverse of Eq.~(\ref{eq:lmaxQFIMSsDicke}) as a function of the number of modes $M$ into which the initial Dicke state $|j,m\rangle$ is split for different $m$ and a total number of $N=100$ atoms. When the mode entanglement limits $1/k$ (gray lines) are crossed from above, genuine $(k+1)$-partite is detected.}
\label{fig:sDickeQFI}
\end{figure}

To compare with the sensitivity that is accessible by an optimal measurement strategy, we employ, as before, the full optimized expression~(\ref{eq:QFIW}). We obtain 
\begin{align}\label{eq:chim2TFSpn}
\boldsymbol{\chi}^{-2}[\hat{\Psi}_{j,m,\mathrm{PN}},\hat{\mathbf{J}}_{\mathbf{r}}] = \dfrac{j^2-m^2}{j}\mathbf{v}\mathbf{v}^T +\mathbf{1},
\end{align}
with $j=N/2$ and $\mathbf{v}=\{\sqrt{p_1}, \sqrt{p_2}, ...\}$. We obtain 
\begin{align}\label{eq:lmaxQFIsDicke}
\lambda_{\max}(\boldsymbol{\chi}^{-2}[\hat{\Psi}_{j,m,\mathrm{PN}},\hat{\mathbf{J}}_{\mathbf{r}}])=\frac{j(j+1)-m^2}{j},    
\end{align}
which indeed coincides with the quantum Fisher information of the Dicke state before splitting for arbitrary rotations in the $xy$-plane~(\ref{eq:QFIDicke}), normalized by the shot-noise level $N=2j$. The resulting sensitivity is shown for comparison in Fig.~\ref{fig:sDicke2} as dashed lines. 

Similarly, we may analyze the mode entanglement using the matrix~(\ref{eq:MSmatrixNL}) and its optimized version~(\ref{eq:QFIMS}). The latter can be compactly expressed as
\begin{align}\label{eq:chim2TFSpnMS}
\boldsymbol{\chi}^{-2}_{\mathrm{MS}}[\hat{\Psi}_{j,m,\mathrm{PN}},\hat{\mathbf{J}}_{\mathbf{r}}] = (j^2-m^2)\mathbf{u}\mathbf{u}^T +\mathbf{F},
\end{align}
where $\mathbf{u}$ is a vector and $\mathbf{F}$ a diagonal matrix with entries
\begin{align}
    u_k&=\sqrt{\frac{p_k}{(j^2-m^2)p_k+j}},\\
    F_k&=\frac{j}{(j^2-m^2)p_k+j}.
\end{align}
We obtain in the case of uniform splitting ratio, i.e., $p_k=1/M$ for all $k$ that 
\begin{align}\label{eq:lmaxQFIMSsDicke}
\lambda_{\max}(\boldsymbol{\chi}^{-2}_{\mathrm{MS}}[\hat{\Psi}_{j,m,\mathrm{PN}},\hat{\mathbf{J}}_{\mathbf{r}}])=\frac{j(j+1)-m^2}{\frac{j^2-m^2}{M}+j}.
\end{align}
In the limit of an infinite number of modes, we obtain again that 
\begin{align}
\lim_{M\to\infty}\lambda_{\max}(\boldsymbol{\chi}^{-2}_{\mathrm{MS}}[\hat{\Psi}_{j,m,\mathrm{PN}},\hat{\mathbf{J}}_{\mathbf{r}}])=\lambda_{\max}(\boldsymbol{\chi}^{-2}[\hat{\Psi}_{j,m,\mathrm{PN}},\hat{\mathbf{J}}_{\mathbf{r}}]),
\end{align}
which is given in Eq.~(\ref{eq:lmaxQFIsDicke}). The mode entanglement detected by the criterion~(\ref{eq:hierarchyMS}) from the quantum Fisher matrix is shown in Fig.~\ref{fig:sDickeQFI}. However, for the chosen local measurement observables, the spin-squeezing matrix~(\ref{eq:MSmatrixNL}) is unable to reveal mode entanglement of split Dicke states.

Summarizing the findings of this Section, we note that if optimal measurements are available, the highly sensitive Dicke states can be converted into an equally sensitive resource for multiparameter estimation through splitting into several spatial modes. Moreover, the splitting generates entanglement among large numbers of modes, which can be detected using metrological entanglement criteria.

Implementing an optimal measurement for spatially distributed non-Gaussian entangled states is, however, more challenging than in the case of Gaussian states. The reason is that the sum of local observables does not correspond to the global optimal observable unless it is linear. Hence, the squeezing matrix of split Dicke states obtained from local, nonlinear measurements describes a multiparameter sensitivity that remains considerably below the ultimate quantum limit. Yet, since the state is pure and the parameters are encoded locally with commuting generators, there exists another measurement strategy that attains the sensitivity described by the quantum Fisher matrix~\cite{Matsumoto2002,PezzePRL2017}.

\section{Application: Nonlocal sensing of a magnetic field gradient}
An application of practical interest is the estimation of magnetic field gradients~\cite{AltenburgPRA2017, ApellanizPRA2018}. Here, we use our results to analyze the sensitivity that can be achieved for this task using split BECs in nonclassical spin states.
In particular, we consider the case of a spin-squeezed BEC split into two modes~\cite{FadelSCIENCE2018} for the estimation of the difference of the magnetic field strength in two spatial positions. In each mode, the local magnetic field leads to a rotation of the spin state due to the Zeeman effect, yielding a parameter-imprinting evolution described by~(\ref{eq:unitary}), where $\theta_k$ depends on the local magnetic field strength and where the direction $\mathbf{r}_k$ can be manipulated by suitable local rotations of the spin state. In the following, we assume that the state is oriented such that the effective rotation axis $\mathbf{r}_k$ corresponds to the local axes of maximal sensitivity that were discussed in Sections~\ref{sec:SSS} and~\ref{sec:STF}.

\begin{figure}[hbt!]
\includegraphics[width=.38\textwidth]{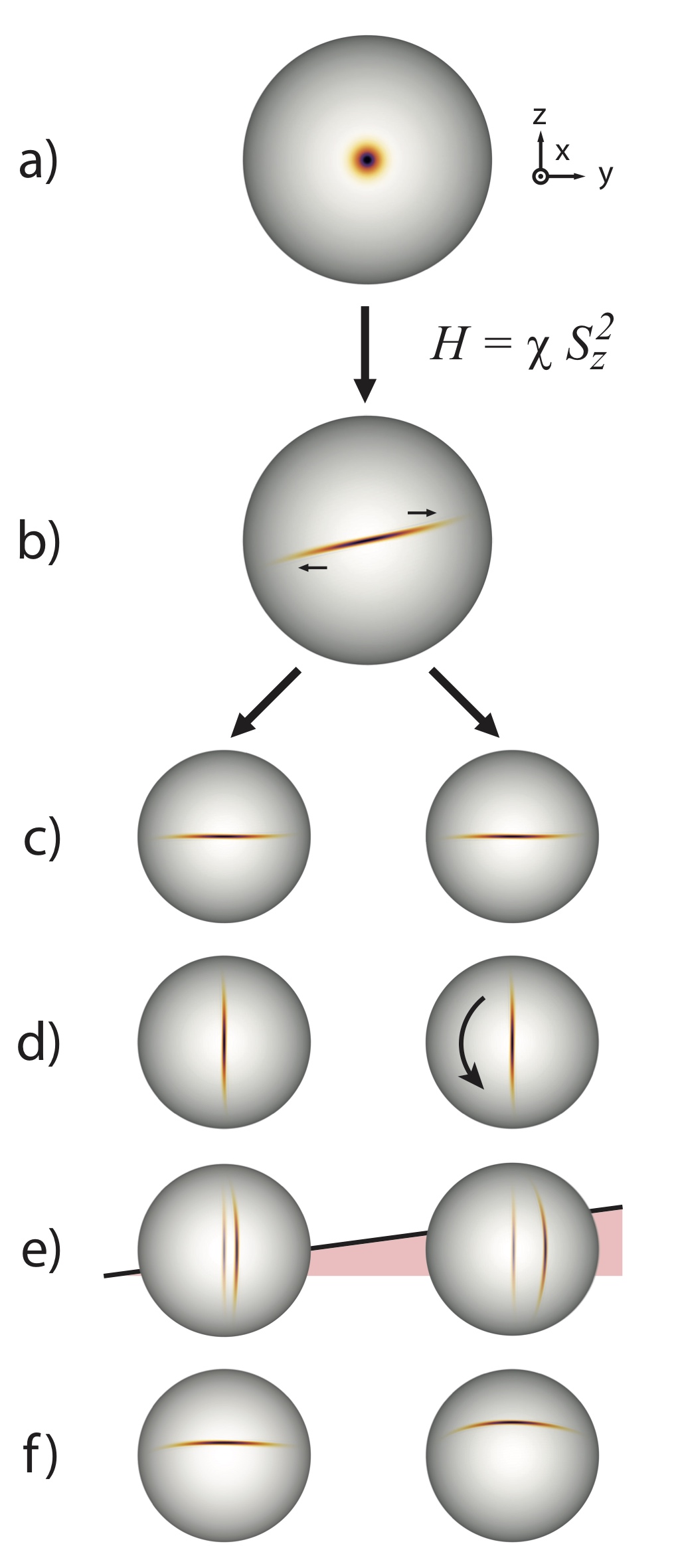}
\caption{\textbf{Experimental protocol for sensing a gradient with a split spin-squeezed state.} Nonclassical correlations are created by exposing a coherent spin state (a) to a nonlinear evolution, leading after a short time to a squeezed spin state (b). Splitting the external degree of freedom into two modes creates a split squeezed state. For the splitting, the state's fluctuations are aligned along the $z$ axis by suitable rotations in order to minimize phase noise (c). To prepare for the Ramsey protocol, the states are rotated such that subsequent phase rotation around the $z$ axis displace the state along its squeezed spin component (d). Moreover, for the estimation of a gradient the second system is rotated $180$ degrees around its mean-spin direction $x$. In the presence of a gradient, the two local spin states experience different rotation angles (e). A final $\pi/2$-pulse around $x$ closes the Ramsey sequence and allows us to estimate the phases from measurements of the relative populations (i.e., the spin $z$-components) in each mode (f).}
\label{fig:3}
\end{figure}

We focus on an estimation of the parameter difference $\theta_A-\theta_B$, which contains information about the magnetic field difference and therefore its gradient. In order to assess the role of the mode entanglement for achieving this measurement sensitivity, we compare our protocol to a local strategy consisting of using the same local states without correlations between the modes. We note that, for the sake of experimental feasibility, we consider a realistic, finite, and fixed amount of squeezing, in contrast to our theoretical analysis of Sec.~\ref{sec:modegain}, where the squeezing of global and local strategies was independently optimized to determine the ultimate limits of each strategy.

As a concrete example, we consider a $^{87}$Rb BEC of $N=1000$ atoms that through OAT dynamics is prepared in a $\xi^2 = -10\,\text{dB}$ spin-squeezed state of the two hyperfine states $\vert F=1, m_F=-1\rangle$ and $\vert F=2, m_F=1\rangle$, Fig.~\ref{fig:3}a,b. By controlling the external trapping potential it is possible to distribute the particles into spatially separated modes \cite{FadelBook}, Fig.~\ref{fig:3}c. During this operation the state can be oriented horizontally (Fig.~\ref{fig:3}c), so that the squeezed quadrature is less affected by phase noise \cite{OckeloenPRL13}. To make a quantitative prediction for the sensitivity, we assume an equal splitting of the atoms into two modes separated by $d=50\,\mu\text{m}$, which is at least a factor 10 larger than the BEC wavefunction size for typical trapping frequencies \cite{FadelSCIENCE2018,FadelBook}. The advantage of using BECs for sensing is in fact that they are extremely localized ensembles, allowing to probe small volumes of space. 

The interferometric (Ramsey) protocol begins with orienting the states vertically, Fig.~\ref{fig:3}d, to maximize the sensitivity to local phase imprinting. In Sec.~\ref{sec:SSSPN} we have seen that, in order to prepare an optimal state for the measurement of the phase difference, it is now convenient to rotate system B's local spin state by $180^\circ$ around the $x$-axis (the mean-spin direction), in order to reverse the sign of the covariance $\mathrm{Cov}(\hat{J}_{\mathbf{s}_A,A},\hat{J}_{\mathbf{s}_B,B})_{\hat{\rho}}$ of the local measurement observables between the two modes. The consequence of this rotation to the spin-squeezing matrix~(\ref{eq:winelandmatrix}) is that the off-diagonal elements acquire a minus sign, while the rest of the elements is unchanged. This maps the linear combination of maximal sensitivity from $(\theta_A+\theta_B)/\sqrt{2}$ to $(\theta_A-\theta_B)/\sqrt{2}$, which is of interest here.

In the presence of a field gradient, the two local states will acquire a different rotation angle depending on their position, see Fig.~\ref{fig:3}e. The interferometic protocol is terminated with a $\pi/2$-pulse around the $x$ axis, Fig.~\ref{fig:3}f which allows to access the local phases by measuring the local population imbalances.

This protocol makes optimal use of the mode entanglement and leads to a sensitivity enhancement that coincides with the squeezing of the atomic ensemble before the splitting (see Sec.~\ref{sec:SSS}), assuming that the splitting process does not introduce additional sources of noise. Since the spin-squeezing matrix quantifies the quantum gain over the shot-noise limit, we obtain the absolute sensitivity by appropriate multiplication with the shot-noise sensitivity, see Sec.~\ref{sec:theory}. For the specific case discussed here, we obtain an uncertainty for the phase difference of $\Delta((\theta_A-\theta_B)/\sqrt{2})=\xi/\sqrt{N}\simeq 3.2\,\text{mrad}$.

The contribution of the mode entanglement can be revealed by treating the two BECs as independent ensembles for comparison. To this end, we study the properties of a reference state $\rho_A\otimes\rho_B$ that has been prepared as the product of the two reduced states of modes $A$ and $B$, respectively. Each subsystem consists of $N_A=N_B=500$ atoms, and the local Wineland spin-squeezing coefficient $\xi_A^2=\xi_B^2 = -2.56\,\text{dB}$ is limited by partition noise and coincides for both modes. 
The squeezing matrix reads $\bs{\xi}^2[\hat{\rho}^A\otimes\hat{\rho}^B,\hat{\mathbf{J}}_{\mathbf{r}},\hat{\mathbf{J}}_{\mathbf{s}}]=\xi_A^2\mathbf{1}_2$.
The degeneracy of this matrix implies that the sensitivity gain is the same for arbitrary normalized linear combinations $\mathbf{n}^T\bm{\theta}=n_A\theta_A+n_B\theta_B$ of the two local phases $\theta_A$ and $\theta_B$ for this local state and reads $\mathbf{n}^T\bs{\xi}^2[\hat{\rho}^A\otimes\hat{\rho}^B,\hat{\mathbf{J}}_{\mathbf{r}},\hat{\mathbf{J}}_{\mathbf{s}}]\mathbf{n}=\xi_A^2$, whenever $n_A^2+n_B^2=1$ (the gradient estimation considered here corresponds to $n_A=-n_B=1/\sqrt{2})$.
Renormalizing the sensitivity gain, as before, with respect to the shot-noise limit, we obtain a sensitivity of
$\Delta((\theta_A-\theta_B)/\sqrt{2})=\xi_A/\sqrt{N_A}\simeq 25\,\text{mrad}$.

\section{Conclusions}
The squeezing matrix represents a practical approach for quantifying multiparameter quantum gain of split squeezed states, and relates the quantum sensitivity advantage to the squeezing of a family of local observables. We have provided exact analytical expressions for the spin-squeezing matrices of nonclassical spin states that are relevant in current experiments with cold and ultra-cold atomic ensembles. Our analysis reveals practical and optimal state preparation and measurement strategies that maximize the multiparameter sensitivity for any linear combination of spatially distributed phase parameters.

For split squeezed states, the collective squeezing in multiparameter measurements coincides with the total squeezing of the spin ensemble before the splitting -- independently of the presence of partition noise in the splitting process. Comparison with the quantum Fisher matrix reveals the optimality of the chosen local measurement strategy as long as the state is Gaussian.

Our framework is applicable to arbitrary pure and mixed quantum states and allows us to include more general, nonlinear measurement observables. An analysis of nonlinear observables on split Dicke states points out the limitations of local measurements for non-Gaussian spin states.

Moreover, we have introduced a way to detect and put quantitative bounds on multimode entanglement directly from information about multiparameter squeezing. This experimentally practical method efficiently detects genuine multimode entanglement of split squeezed states.

Finally, we have studied the performance of these states for gradient sensing with realistic experimental parameters, and illustrated the metrological advantage provided by mode entanglement.

Our results outline concrete strategies for harnessing the nonclassical features of spatially split squeezed states for quantum-enhanced multiparameter measurements in an optimal way. These results provide relevant guidance for ongoing experiments with Bose-Einstein condensates.

In future works, it would be interesting to investigate how the spin-squeezing matrix could give a quantification of entanglement through a connection with entanglement monotones \cite{HuberReview,FadelPRL21}, and the metrological advantage provided by correlations stronger than entanglement \cite{Frowis,Yadin21,Guo21,Meng21}.

\section{Acknowledgments}

MF was supported by the Swiss National Science Foundation, and by The Branco Weiss Fellowship -- Society in Science, administered by the ETH Z\"{u}rich.
TB was supported by the National Natural Science Foundation of China (62071301); State Council of the People’s Republic of China (D1210036A); NSFC Research Fund for International Young Scientists (11850410426); NYU-ECNU Institute of Physics at NYU Shanghai; the Science and Technology Commission of Shanghai Municipality (19XD1423000); the China Science and Technology Exchange Center (NGA-16-001); the NYU Shanghai Boost Fund. BY has received funding from the European Union's Framework Programme for Research and Innovation Horizon 2020 (2014-2020) under the Marie Sk\l odowska-Curie Grant Agreement No.\ 945422. MG was funded by MCIN / AEI for the project PID2020-115761RJ-I00 and by a fellowship from ``la Caixa" Foundation (ID 100010434) and by the European Union's Horizon 2020 Research and Innovation Programme under the Marie Sk\l{}odowska-Curie grant agreement No. 847648, fellowship code LCF/BQ/PI21/11830025.

\clearpage
\newpage

\begin{widetext}
\appendix

\section*{Supplementary material}

Here we show the detailed calculations for some of the results presented in the paper.

\section{$k$-producibility bound} \label{ap:k-prod-witness}

In this appendix we prove the bound~(\ref{eq:ksepMS}). The result follows from
\begin{align} \label{eq:qfi_bound}
	F_Q[\hat{\rho}_{k-\mathrm{prod}}, \sum_{i=1}^M \hat{H}_{i}] \leq 4k \sum_{i=1}^M (\Delta \hat{H}_{i})^2,
\end{align}
which holds for any $k$-producible state $\hat{\rho}_{k-\mathrm{prod}}$ for arbitrary local observables $\hat{H}_i$ in the modes $i=1,\dots,M$. Here, $F_Q[\hat{\rho}, \hat{H}]$ is the quantum Fisher information, which expresses the sensitivity of the state $\hat{\rho}$ to unitary transformations generated by $\hat{H}$~\cite{PezzeRMP2018}. 

\begin{proof}
We break the proof into several stages. Firstly, we find the maximal quantum Fisher information allowing for any amount of entanglement between $M$ subsystems. Consider an arbitrary set of local observables $\hat{\mathbf{H}}=(\hat{H}_1,\dots,\hat{H}_M)^T$. First note that $F_Q[\hat{\rho},\sum_{i=1}^M \hat{H}_i] \leq 4 (\Delta\sum_{i=1}^M \hat{H}_i)^2_{\hat{\rho}}$, and then
\begin{align}
	(\Delta\sum_{i=1}^M \hat{H}_i)^2_{\hat{\rho}}
		& = \sum_{i,j=1}^M \mathrm{Cov}(\hat{H}_i,\hat{H}_j)_{\hat{\rho}} \nonumber \\
		& \leq \sum_{i,j=1}^M (\Delta \hat{H}_i)_{\hat{\rho}} (\Delta \hat{H}_j)_{\hat{\rho}} \nonumber \\
		& = \left( \sum_{i=1}^M (\Delta \hat{H}_i)_{\hat{\rho}} \right)^2,
\end{align}
where the inequality follows from $\mathrm{Cov}(\hat{H}_i,\hat{H}_j)_{\hat{\rho}}^2 \leq (\Delta\hat{H}_i)_{\hat{\rho}}^2(\Delta\hat{H}_j)_{\hat{\rho}}^2$. The Cauchy-Schwarz inequality implies $[\sum_{i=1}^M (\Delta\hat{H}_i)_{\hat{\rho}}/M]^2 \leq \sum_{i=1}^M (\Delta\hat{H}_i)_{\hat{\rho}}^2/M$, so
\begin{align} \label{eq:qfi_maximal}
	F_Q[\hat{\rho},\sum_{i=1}^M \hat{H}_i] \leq 4 (\Delta\sum_{i=1}^M \hat{H}_i)^2_{\hat{\rho}} \leq 4M^2 \sum_{i=1}^M \frac{1}{M} (\Delta\hat{H})_{\hat{\rho}}^2 
		 = 4M \sum_{i=1}^M (\Delta\hat{H})_{\hat{\rho}}^2.
\end{align}
Next, take any product state $\hat{\rho}=\hat{\rho}_1 \otimes \hat{\rho}_2$, in which $\hat{\rho}_1$ and $\hat{\rho}_2$ exist on $M$ and $M'$ subsystems, respectively. Then additivity of the quantum Fisher information, followed by \eqref{eq:qfi_maximal}, gives
\begin{align}
	F_Q[\hat{\rho}_1 \otimes \hat{\rho}_2, \sum_{i=1}^{M+M'} \hat{H}_i] & = F_Q[\hat{\rho}_1,\sum_{i=1}^M \hat{H}_i] + F_Q[\hat{\rho}_2, \sum_{i=M+1}^{M+M'} \hat{H}_i] \nonumber \\
	& \leq 4M \sum_{i=1}^M (\Delta\hat{H}_i)_{\hat{\rho}}^2 + 4M' \sum_{i=M+1}^{M+M'} (\Delta\hat{H}_i)_{\hat{\rho}}^2 \nonumber \\
	& \leq 4\max\{M,M'\} \sum_{i=1}^{M+M'} (\Delta\hat{H}_i)_{\hat{\rho}}^2.
\end{align}
It follows straightforwardly that, if we divide $M$ subsystems into $b$ blocks whose size is each no greater than $k$, then any product state $\hat{\rho} = \bigotimes_{\alpha=1}^b \hat{\rho}_\alpha$ with respect to this structure satisfies
\begin{align} \label{eq:qfi_block}
	F_Q[\bigotimes_{\alpha=1}^b \hat{\rho}_\alpha, \sum_{i=1}^M \hat{H}_i] \leq 4k \sum_{i=1}^M (\Delta\hat{H}_i)_{\hat{\rho}}^2.
\end{align}
Finally, a $k$-producible state $\hat{\rho}_{k-\mathrm{prod}}$ is by definition a mixture of such product states $\hat{\rho}^{(j)}$, each of which has entangled blocks of size no greater than $k$ but may have different partition structures. Convexity of the quantum Fisher information and concavity of the variance then implies
\begin{align}
    F_Q[\hat{\rho}_{k-\mathrm{prod}}, \sum_{i=1}^M \hat{H}_i] & \leq \sum_j p_j F_q[\hat{\rho}^{(j)}, \sum_{i=1}^M \hat{H}_i] \nonumber \\
    & \leq 4k \sum_j p_j \sum_{i=1}^M (\Delta \hat{H}_i)^2_{\hat{\rho}^{(j)}} \nonumber \\
    & \leq 4k \sum_{i=1}^M (\Delta \hat{H}_i)^2_{\hat{\rho}},
\end{align}
using \eqref{eq:qfi_block} in the second line.
\end{proof}

Now consider a linear combination of local observables $\hat{H}_{\mathbf{n}}=\mathbf{n}^T\hat{\mathbf{H}}=\sum_{i=1}^Mn_i\hat{H}_i$. Using $F_Q[\hat{\rho},\hat{H}_{\mathbf{n}}]=\mathbf{n}^T\mathbf{F}_Q[\hat{\rho},\hat{\mathbf{H}}]\mathbf{n}$, where $\mathbf{F}_Q[\hat{\rho},\hat{\mathbf{H}}]$ is the quantum Fisher matrix~\cite{GessnerPRL2018} we can rewrite~(\ref{eq:qfi_bound}) as
\begin{align}
    \mathbf{n}^T\mathbf{F}_Q[\hat{\rho}_{k-\mathrm{prod}},\hat{\mathbf{H}}]\mathbf{n}\leq 4k \mathbf{n}^T \mathrm{diag}((\Delta\hat{H}_{1})_{\hat{\rho}_{k-\mathrm{prod}}}^2,\dots,(\Delta\hat{H}_{M})_{\hat{\rho}_{k-\mathrm{prod}}}^2)\mathbf{n}.
\end{align}
This statement holds for arbitrary $\mathbf{n}$ and can therefore be stated as a matrix inequality
\begin{align}\label{eq:upperFQ}
    \mathbf{F}_Q[\hat{\rho}_{k-\mathrm{prod}},\hat{\mathbf{H}}]\leq 4k\mathrm{diag}((\Delta\hat{H}_{1})_{\hat{\rho}_{k-\mathrm{prod}}}^2,\dots,(\Delta\hat{H}_{M})_{\hat{\rho}_{k-\mathrm{prod}}}^2).
\end{align}
Moreover, the moment matrix represents a lower bound on the quantum Fisher matrix~\cite{GessnerNATCOMMUN2020}, i.e., $\mathbf{M}[\hat{\rho},\hat{\mathbf{H}},\hat{\mathbf{X}}]\leq \mathbf{F}_Q[\hat{\rho},\hat{\mathbf{H}}]$ for arbitrary states $\hat{\rho}$ and sets of local observables $\hat{\mathbf{H}},\hat{\mathbf{X}}$. Consequently, the upper bound~(\ref{eq:upperFQ}) also applies to the moment matrix of $k$-separable states, and formulating in terms of local spin observables leads to
\begin{align}
\mathbf{M}[\hat{\rho}_{k-\mathrm{prod}},\hat{\mathbf{J}}_{\mathbf{r}},\hat{\mathbf{J}}_{\mathbf{s}}]\leq 4k\mathrm{diag}((\Delta\hat{J}_{\mathbf{r}_1,1})_{\hat{\rho}_{k-\mathrm{prod}}}^2,\dots,(\Delta\hat{J}_{\mathbf{r}_M,M})_{\hat{\rho}_{k-\mathrm{prod}}}^2).
\end{align}
From this we can derive the limit on the mode-separability spin-squeezing matrix~(\ref{eq:MSmatrix}) following analogous steps as for the derivation of Eqs.~(\ref{eq:sqz_cond}) and~(\ref{eq:sqz_condMS}). We finally obtain the result~(\ref{eq:ksepMS}).

\section{Split Spin-Squeezed state, with partition noise} \label{app:SSSwithPN}
We consider the state $\vert \Psi(\mu) \rangle = e^{- i \frac{\mu}{2} J_z^2} |N/2\rangle_x$ that is generated from the OAT dynamics. By applying a spatial beam-splitter transformation, the correlated spins are distributed into $M$ modes with a ratio determined by the probability distribution $p_1,\dots,p_M$, so that on average we find $N_k=p_k N$ particles in mode $k$. For any modes $1\leq k,l\leq M$ and local collective angular momentum observables $\mathbf{u}=(u_x,u_y,u_z)^T$ and $\mathbf{v}=(v_x,v_y,v_z)^T$, we obtain the expectation values 
\begin{align}
\avg{\hat{J}_{\mathbf{u},k}} &= p_k \dfrac{N}{2} u_x \cos\left(\dfrac{\mu}{2}\right)^{N-1}\\
\avg{(\hat{J}_{\mathbf{u},k})^2} &= p_k \dfrac{N}{4} + p_k^2 \dfrac{N(N-1)}{8} \left( (u_x^2+u_y^2) + (u_x^2-u_y^2) \cos\left(\mu \right)^{N-2} + 4 u_y u_z \cos\left(\dfrac{\mu}{2}\right)^{N-2} \sin\left(\dfrac{\mu}{2}\right)\right)\\
\avg{\hat{J}_{\mathbf{u},k} \hat{J}_{\mathbf{v},l}} &= p_k p_l \dfrac{N(N-1)}{8}  \left( (u_x v_x + u_y v_y) + (u_x v_x - u_y v_y) \cos\left(\mu \right)^{N-2} + 2 (u_y v_z + u_z v_y) \cos\left(\dfrac{\mu}{2}\right)^{N-2} \sin\left(\dfrac{\mu}{2}\right)\right).
\end{align}
We observe that in each mode, the resulting state is polarized along the $x$, and it shows along a squeezing direction $\mathbf{s}=(0,-\sin(\theta_s),\cos(\theta_s))^T$ on the $zy$-plane a smaller variance than the spin-coherent state, originating from the entanglement created by the nonlinear evolution. The angle $\theta_s$ is found analytically by diagonalizing the covariance matrix to be
\begin{equation}\label{eq:SqAngle}
    \theta_s = \dfrac{1}{2} \arctan \left( \dfrac{4 \sin (\frac{\mu}{2})\cos^{N-2}(\frac{\mu}{2})}{1-\cos^{N-2}(\mu)} \right) \;,
\end{equation}
and it represents the angle measured counterclockwise from the $y$-axis to the direction with maximal variance, i.e., the anti-squeezing direction $\mathbf{r}=(0,\cos(\theta_s),\sin(\theta_s))^T$. Interestingly, this angle is the same for all modes, i.e., it does not depend on the mode index $k$. It is easy to check that the local squeezing direction coincides with the squeezing direction for the state before splitting. This is expected from the fact that the beam-splitter transformation does not act on the internal degrees of freedom, \ie it does not rotate the spin state.

\section{Split Spin-Squeezed state, without partition noise}
\label{app:SSSwithoutPN}

$N=N_A+N_B+...$ is the total number of particles before the splitting.

\begin{equation}
\avg{\hat{J}_{\mathbf{u},k}} = \dfrac{N_k}{2} u_x \cos\left(\dfrac{\mu}{2}\right)^{N-1}
\end{equation}

\begin{equation}
\avg{(\hat{J}_{\mathbf{u},k})^2} = \dfrac{N_k}{4} u_z^2 + \dfrac{N_k}{8} \left( (N_k+1)(u_x^2+u_y^2) + (N_k-1)\left( (u_x^2-u_y^2) \cos\left(\mu \right)^{N-2} + 4 u_y u_z \cos\left(\dfrac{\mu}{2}\right)^{N-2} \sin\left(\dfrac{\mu}{2}\right)\right) \right)
\end{equation}

\begin{equation}
\avg{\hat{J}_{\mathbf{u},k} \hat{J}_{\mathbf{v},l}} = \dfrac{N_k N_l}{8}  \left( (u_x v_x + u_y v_y) + (u_x v_x - u_y v_y) \cos\left(\mu \right)^{N-2} + 2 (u_y v_z + u_z v_y) \cos\left(\dfrac{\mu}{2}\right)^{N-2} \sin\left(\dfrac{\mu}{2}\right)\right)
\end{equation}

\section{Split Fock state, with partition noise}
Consider the vectors $\mathbf{u}=(u_x,u_y,u_z)$, $\mathbf{v}=(v_x,v_y,v_z)$, and $m=-j...j$.

\begin{equation}
\avg{\hat{J}_{\mathbf{u},k}} = p_k m u_z 
\end{equation}

\begin{equation}
\avg{(\hat{J}_{\mathbf{u},k})^2} = \dfrac{p_k}{2} \left( (j + (j^2-m^2) p_k)(u_x^2+u_y^2) + ( j - j p_k + 2 m^2 p_k) u_z^2 \right)
\end{equation}

\begin{equation}
\avg{\hat{J}_{\mathbf{u},k} \hat{J}_{\mathbf{v},l}} = \dfrac{j^2-m^2}{2} p_k p_l \left( u_x v_x + u_y v_y \right) + \left( m^2 - \dfrac{j}{2} \right) p_k p_l u_z v_z
\end{equation}

\section{Single-mode Dicke state}\label{app:DickeSingleMode}

Let us consider the basis of operators
\begin{equation}
H=\left(  \hat{J}_x, \hat{J}_y, \hat{J}_z, \left\lbrace  \hat{J}_x, \hat{J}_z \right\rbrace /2, \left\lbrace  \hat{J}_x, \hat{J}_y \right\rbrace /2 , \left\lbrace  \hat{J}_x, \hat{J}_z \right\rbrace /2 , \hat{J}_x^2, \hat{J}_y^2, \hat{J}_z^2\right)  \end{equation}

Relevant covariances are

\begin{equation}
\text{Cov}(\hat{J}_x,\hat{J}_x) = \text{Cov}(\hat{J}_y,\hat{J}_y) = \dfrac{1}{2} \left( j (j+1) - m^2 \right)
\end{equation}

\begin{equation}
\text{Cov}(\hat{J}_z,\hat{J}_z) = \text{Cov}(\hat{J}_{i},\hat{J}_{j\neq i}) = 0
\end{equation}

\begin{equation}
\text{Cov}(\hat{J}_{i},\hat{J}_{j}^2) = 0
\end{equation}

\begin{equation}
\text{Cov}(\hat{J}_x^2,\hat{J}_x^2) = \text{Cov}(\hat{J}_y^2,\hat{J}_y^2) = \dfrac{1}{8} \left( m^2(m^2+5) + j(j+1)(j(j+1)-2(m^2+1))  \right)
\end{equation}

\begin{equation}
\text{Cov}(\hat{J}_x^2,\hat{J}_y^2) = -\dfrac{1}{8} \left( m^2(m^2+5) + j(j+1)(j(j+1)-2(m^2+1))  \right)
\end{equation}

\begin{equation}
\text{Cov}(\hat{J}_x,\{ \hat{J}_x, \hat{J}_z \}/2) = \text{Cov}(\hat{J}_y,\{ \hat{J}_y, \hat{J}_z \}/2) = \dfrac{m}{4}\left( 2j(j+1)-(2m^2+1) \right)
\end{equation}

\begin{equation}
\text{Cov}(\hat{J}_{i}^2, \{ \hat{J}_{j}, \hat{J}_{k} \}/2 ) = 0
\end{equation}

\begin{equation}
\text{Cov}(\{ \hat{J}_x, \hat{J}_z \}/2,\{ \hat{J}_x, \hat{J}_z \}/2) = \text{Cov}(\{ \hat{J}_y, \hat{J}_z \}/2,\{ \hat{J}_y, \hat{J}_z \}/2) = \dfrac{1}{8} \left(j(j+1) + (4j(j+1)-5)m^2 - 4m^4  \right)
\end{equation}

\begin{equation}
\text{Cov}( \{ \hat{J}_x, \hat{J}_y \}/2, \{ \hat{J}_x, \hat{J}_y \}/2) = \dfrac{1}{8} \left( m^2(m^2+5) + j(j+1)(j(j+1)-2(m^2+1)) \right)
\end{equation}

For the commutator matrix, we consider only commutators between a linear spin observable, and an element of the set $H$.
The only non-zero commutators turn out to be 
\begin{equation}
-i\langle [\hat{J}_x,\hat{J}_y]\rangle=m 
\end{equation}

\begin{equation}
-i\langle [\hat{J}_x,\{ \hat{J}_y, \hat{J}_z \}/2]\rangle = i\langle [\hat{J}_y,\{ \hat{J}_x, \hat{J}_z \}/2]\rangle = - \dfrac{1}{2}(j(j+1)-3m^2)
\end{equation}

\section{Split Dicke state with partition noise}\label{app:splitDicke}

We consider the sets of nonlinear operators
\begin{equation}
	H_k=\left(  \hat{J}_{x,k}, \hat{J}_{y,k}, \left\lbrace  \hat{J}_{x,k}, \hat{J}_{z,k} \right\rbrace /2, \left\lbrace  \hat{J}_{y,k}, \hat{J}_{z,k} \right\rbrace /2 \right) 
\end{equation}

The only non-zero expectation values of a commutator between $\hat{J}_{y,k}$ and an element of $H_k$ are
\begin{equation}
\langle [\hat{J}_{x,k},\hat{J}_{y,k}]\rangle = i m p_k 
\end{equation}

\begin{equation}
\langle\left[\hat{J}_{y,k},\{\hat{J}_{x,k},\hat{J}_{z,k}\}/2 \right]\rangle =  \dfrac{i}{2} \left( j(j+1) - 3 m^2 \right)p_{k}^{2}
\end{equation}

Non-zero covariances are
\begin{equation}
\text{Cov}(\hat{J}_{x,k},\hat{J}_{x,k}) = \text{Cov}(\hat{J}_{y,k},\hat{J}_{y,k}) = \dfrac{1}{2} p_k (j + (j^2 - m^2) p_k)
\end{equation}

\begin{equation}
\text{Cov}(\hat{J}_{x,k},\{\hat{J}_{x,k},\hat{J}_{z,k}\}/2) = \text{Cov}(\hat{J}_{y,k},\{\hat{J}_{y,k},\hat{J}_{z,k}\}/2) = \dfrac{1}{2} p_k^2 \left( \dfrac{1}{2} (2j-1)m + m(j^2-m^2)p_k \right)
\end{equation}

\begin{align}
&\text{Cov}(\{\hat{J}_{x,k},\hat{J}_{z,k}\}/2,\{\hat{J}_{x,k},\hat{J}_{z,k}\}/2) = \text{Cov}(\{\hat{J}_{y,k},\hat{J}_{z,k}\}/2,\{\hat{J}_{y,k},\hat{J}_{z,k}\}/2) = \\ 
&\qquad = \dfrac{1}{8} p_k^2 (j (3 j - 1) - m^2 + 2 (j-1) ((j-1) j + m^2) p_k -  2 (j - m) (j + m) (j - 2 m^2 - 1) p_k^2)
\end{align}

\begin{equation}
\text{Cov}(\hat{J}_{x,k},\hat{J}_{x,l}) = \text{Cov}(\hat{J}_{y,k},\hat{J}_{y,l}) = \dfrac{1}{2} p_k p_l (j^2 - m^2)
\end{equation}

\begin{equation}
\text{Cov}(\hat{J}_{x,k},\{\hat{J}_{y,l},\hat{J}_{z,l}\}/2) = \text{Cov}(\hat{J}_{y,k},\{\hat{J}_{y,l},\hat{J}_{z,l}\}/2) = \dfrac{1}{2} p_k p_l^2 m (j^2 - m^2)
\end{equation}

\begin{equation}
\text{Cov}(\{\hat{J}_{x,k},\hat{J}_{z,k}\}/2,\hat{J}_{x,l}) = \text{Cov}(\{\hat{J}_{y,k},\hat{J}_{z,k}\}/2,\hat{J}_{y,l}) = \dfrac{1}{2} p_k^2 p_l m (j^2 - m^2)
\end{equation}

\begin{equation}
\text{Cov}(\{\hat{J}_{x,k},\hat{J}_{z,k}\}/2,\{\hat{J}_{x,l},\hat{J}_{z,l}\}/2) = \text{Cov}(\{\hat{J}_{y,k},\hat{J}_{z,k}\}/2,\{\hat{J}_{y,l},\hat{J}_{z,l}\}/2) = -\dfrac{1}{4} p_k^2 p_l^2 (j - m) (j + m) (j - 2 m^2 - 1) 
\end{equation}

\end{widetext}

\end{document}